\documentclass[prx,eqsecnum,twocolumn,aps,showpacs,amsmath,amssymb,floatfix,superscriptaddress]{revtex4}
\newcommand{\be}{\begin{equation}}
\newcommand{\ee}{\end{equation}}
\newcommand{\bea}{\begin{eqnarray}}
\newcommand{\eea}{\end{eqnarray}}

\usepackage{color}
\usepackage{graphicx}
\usepackage{dcolumn}
\usepackage{bm}
\usepackage{array}
\usepackage{float}
\usepackage{supertabular}
\usepackage{longtable}
\usepackage{mathrsfs}
\usepackage{txfonts}
\usepackage{wasysym}
\usepackage[normalem]{ulem}

\begin{document}
\title{Phase Diagram of the Interacting Majorana Chain Model}
\author{Armin Rahmani}
\affiliation{Department of Physics and Astronomy and 
Quantum Matter Institute, University of British Columbia, Vancouver, British Columbia, Canada V6T 1Z4}

\author{Xiaoyu Zhu}
\affiliation{Department of Physics and Astronomy and 
Quantum Matter Institute, University of British Columbia, Vancouver, British Columbia, Canada V6T 1Z4}
\affiliation{National Laboratory of Solid State Microstructures and Department of Physics, Nanjing University - Nanjing 210093, China}

\author{Marcel Franz}
\affiliation{Department of Physics and Astronomy and 
Quantum Matter Institute, University of British Columbia, Vancouver, British Columbia, Canada V6T 1Z4}

\author{Ian Affleck}
\affiliation{Department of Physics and Astronomy and 
Quantum Matter Institute, University of British Columbia, Vancouver, British Columbia, Canada V6T 1Z4}
\date{\today}
\begin{abstract}
The  Hubbard chain and  spinless fermion chain are paradigms of 
strongly correlated systems, very well understood using Bethe ansatz, Density Matrix Renormalization Group (DMRG) and field theory/renormalization group (RG) methods.  
They have been applied to  one-dimensional materials and have provided important insights for understanding  higher dimensional cases. 
Recently, a new interacting fermion model has been introduced, with possible applications to topological materials. 
It has  a single Majorana fermion operator on each 
lattice site and interactions with the shortest possible range that involve 4 sites. We present a thorough analysis of the phase diagram 
of this model in one dimension using field theory/RG and DMRG methods.  It includes a gapped supersymmetric region and a novel 
gapless phase with coexisting Luttinger liquid and Ising degrees of freedom. In addition to a first order transition, three critical points occur: tricritical Ising, 
Lifshitz and a novel generalization of the commensurate-incommensurate transition. We also survey various gapped phases of the system that arise when the translation symmetry is broken by dimerization and find both trivial and topological phases with 0, 1 and 2 Majorana zero modes bound to the edges of the chain with open boundary conditions.
\end{abstract}
\maketitle

\begin{figure}[t]
\centerline{\includegraphics[width=80mm]{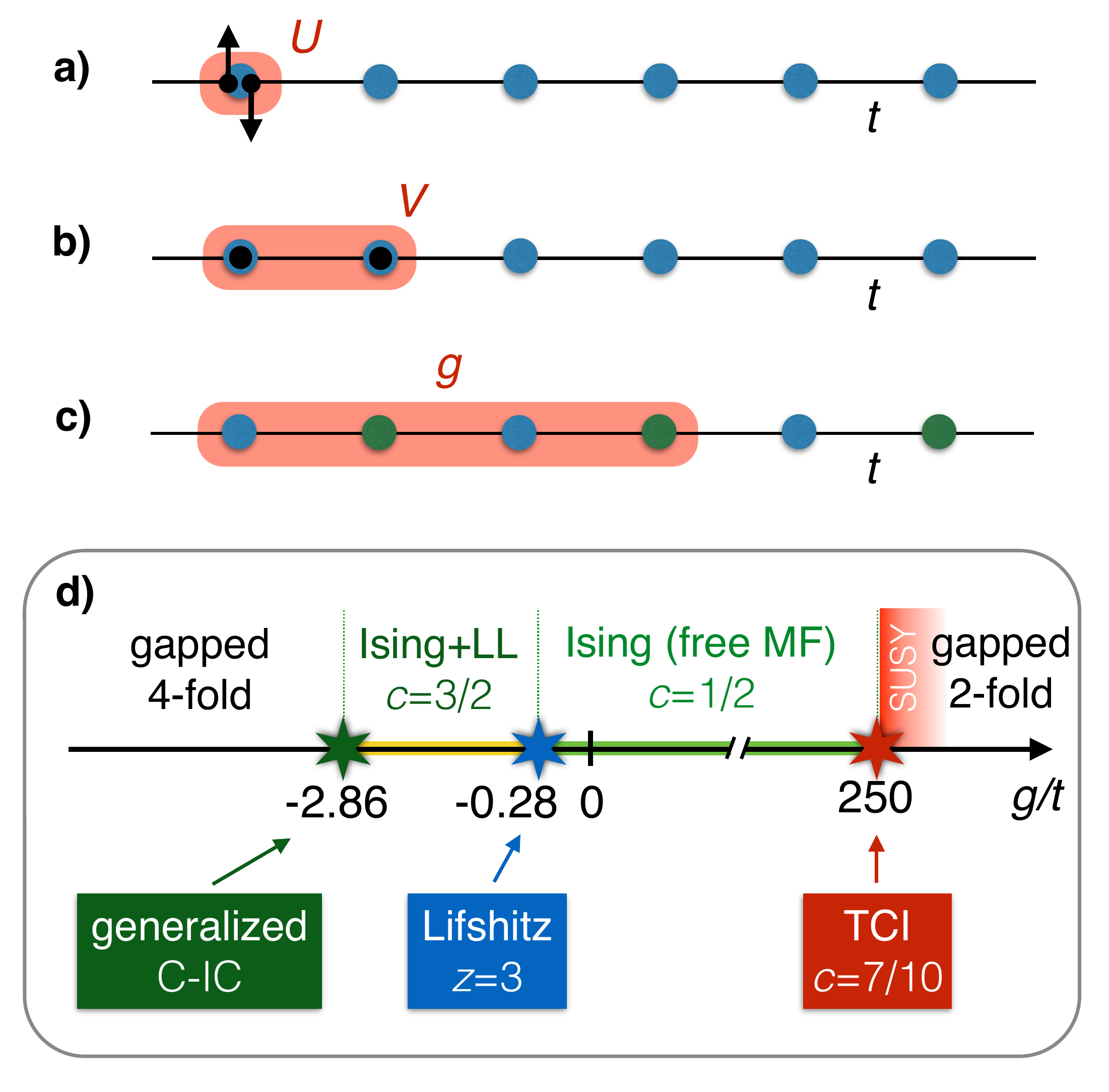}}
\caption{(a) The Hubbard chain with on-site interactions. (b) The spinless Dirac chain with nearest-neighbor interactions. (c) The most local Majorana chain with 4-site interactions. (d) The phase diagram of the Majorana chain with the Hamiltonian~\eqref{eq:hamil} as a function of $g/t$. Setting $t=1$, it consists of four phases: a 4-fold degenerate gapped phase separated by a generalized commensurate-incommensurate (C-IC) transition at $g\approx -2.86$ from a critical phase with central charge $c=3/2$ comprised of a critical Ising and a decoupled Luttinger liquid (LL). The Ising+LL phase is, in turn, separated from a critical Ising phase by a Lifshitz critical point with dynamical exponent $z=3$ at $g\approx -0.28$. For positive $g$, we have another transition from the Ising phase to a doubly degenerate supersymmetric gapped phase at $g\approx 250$ with the phase transition described by the tricritical Ising (TCI) conformal field theory (CFT) with central charge $c=7/10$.   }
\label{phaseD}
\end{figure}

\section{Introduction}
Models of strongly correlated electrons have provided outstanding challenges to condensed matter theory for many decades. These models are expected to describe the physics of important materials such as high temperature superconductors, complex oxides and quantum magnets. While exact or controlled analytical and numerical solutions are rare for the two and three dimensional cases, great understanding has been achieved for the one-dimensional (1D) versions of these models.  
This is true despite the fact that quantum fluctuations are generally enhanced in 1D, leading to the break-down of Fermi liquid and mean-field theories.  The remarkable progress may be ascribed to a number of theoretical techniques: 
the Bethe ansatz, numerical methods, notably the density matrix renormalization group (DMRG)~\cite{White1992}, and field theory methods based on bosonization and the renormalization group (RG).

The canonical example of correlated electrons, namely, the Hubbard chain depicted in Fig.\ \ref{phaseD}(a), has interaction between electrons with opposite spin on the same site, 
\be\label{h0}
H=-t\sum_{j\sigma}(c^\dagger_{j,\sigma}c_{j+1,\sigma}+{\rm H.c.})+U\sum_{j}\hat n_{j,\uparrow}\hat n_{j,\downarrow}.
\ee
It has been well studied by all these methods and its phase diagram is understood in great detail for both 
signs of the on-site interaction $U$  and for general values of the chemical potential $\mu$ (which couples to the total density as $-\mu \sum_{j\sigma} \hat n_{j,\sigma}$). (For a review see~\cite{Essler2005}.)
 In Eq.\ (\ref{h0})  $c_{j\sigma}$ annihilates an electron on site $j$ with spin $\sigma=\uparrow,\downarrow$ and $\hat n_{j\sigma}=c^\dagger_{j\sigma}c_{j\sigma}$. An even simpler version of the Hubbard chain for spinless fermions, which we henceforth refer to as the Dirac chain, is described by the  Hamiltonian
\be H=\sum_j\left[ -t(c^\dagger_jc_{j+1}+{\rm H.c.})+V(\hat n_j-1/2)(\hat n_{j+1}-1/2)\right].\label{Dirac}\ee
The model~\eqref{Dirac} is illustrated in Fig.\ \ref{phaseD}(b). It has also 
been very well studied by all the methods mentioned above~\cite{Bethe1931, Luther1975} for general chemical potentials.
In both models Coulomb interactions are treated as being highly screened, with strictly on-site interactions in the spinful Hubbard case 
and nearest-neighbor interactions, the shortest range possible due to the Pauli exclusion principle, for the spinless case (note that $\hat n^2_j=\hat n_j$).  
The spinless fermion model is equivalent to the XXZ spin-1/2 chain, providing important experimental realizations.

There is a third very natural 1D model which reduces the number of degrees of freedom per site by another factor of 1/2 compared to the spinless fermion chain. The fermions in Eq.~\eqref{Dirac} are complex (Dirac) fermions with $c_j\neq c^\dagger_j$. The analog of Eq.~\eqref{Dirac} for real (Majorana) fermion operators $\gamma_j$, obeying
\be \gamma_j^\dagger =\gamma_j,\ \  \{\gamma_j,\gamma_{i}\}=2\delta_{j,i},\ee
can be written as
\be\label{eq:hamil} H=\sum_j\left[it\gamma_j\gamma_{j+1}+g\gamma_j\gamma_{j+1}\gamma_{j+2}\gamma_{j+3}\right].\ee
The Hamiltonian again has nearest-neighbor hopping and interactions of the shortest possible range: Since $\gamma_j^2=1$, the shortest range nontrivial interaction spans four consecutive sites as shown in Fig.~\ref{phaseD}(c). In this case, no chemical potential term is possible 
so the model only has one dimensionless parameter $g/t$. Surprisingly, this third canonical model of interacting fermions has remained relatively unexplored~\cite{Hassler2012,Terhal2012,Kells2014}. This is partly due to the fact that Majorana fermions are yet to be observed as elementary particles of nature. However, Majorana zero modes (a special type of Majorana fermion occurring at exactly zero energy),   are predicted to occur as collective degrees of freedom in several condensed matter systems~\cite{Read2000,Kitaev2001, Stern2008, Nayak2008,Fu2008,Lutchyn2010,Oreg2010,Alicea2012,Beenakker2012,Elliott2015}, with remarkable experimental progress reported in recent years~\cite{Mourik2012,Das2012,Deng2012,Rokhinson2012,Finck2013,Hart2014,Nadj-Perge2014}.

The model defined by Eq.\ (\ref{eq:hamil}) above has been proposed to describe the low-energy physics of a 1D vortex 
lattice formed in a superconducting film on the surface of a strong topological insulator (STI) \cite{Chiu2015,Chiu2015b,Pikulin2015}. A Majorana zero mode (MZM) exists in the core of 
each vortex~\cite{Fu2008}. The tunneling term in Eq.~\eqref{eq:hamil} makes these Majorana modes dispersive, moving their energy away from zero (corresponding to an isolated MZM). By tuning the chemical potential of the STI, it is possible to tune the hopping parameter $t$ to zero (due to a chiral symmetry~\cite{Teo2010,Chiu2015,Liu2015}), providing access to the strong coupling regime $|g/t|\gg 1$.  It was shown that ordinary Coulomb interactions in the STI lead to $g<0$. 
However, the $g>0$ case might also be relevant due to the effectively attractive electron-electron interactions leading to superconductivity.

Unlike the Hubbard or spinless Dirac fermion chain,
the Majorana chain has no continuous symmetries and is not Bethe ansatz integrable, as far as we know. Also unlike the Hubbard 
and Dirac chains, the Majorana chain has a nontrivial ground state in the strong coupling limit, $t=0$.   To study the model numerically, and understand its behavior in certain limits, it is convenient to construct  Dirac fermion operators from pairs of neighboring 
MZMs. Combining, e.g., sites $2j$ with $2j+1$ into
\be
c_j\equiv (\gamma_{2j}+i\gamma_{2j+1})/2,\label{cdef}\ee
the Hamiltonian takes the form
\be 
\begin{split} 
H=&\sum_j\left\{ t\left[\hat p_j-(c^\dagger_j-c_j)(c^\dagger_{j+1}+c_{j+1})\right]\right.\\&+g\left.\left[-\hat p_j\hat p_{j+1}+(c^\dagger_{j}-c_{j})\hat p_{j+1}(c^\dagger_{j+2}+c_{j+2})\right]\right\},\label{HDir}
\end{split}
 \ee
 where $\hat p_j$ is shorthand for $2\hat n_j-1$.
Note that the hopping term turns into a combination of a chemical potential, nearest-neighbor hopping and nearest-neighbor pairing terms, 
whereas the interaction term turns into a combination of nearest-neighbor interactions and second-neighbor hopping and pairing with amplitude 
depending on the filling at the central site. As anticipated, negative $g$ corresponds to a repulsive interaction term. Similar to the model of Eq.~\eqref{Dirac}, the Majorana chain model can also be 
exactly mapped into a spin Hamiltonian by a Jordan-Wigner transformation $\sigma_j^z=\hat p_j$ and $\sigma_j^+=e^{i\pi\sum_{k<j} \hat n_k}c^\dagger_j$, which leads to 
 \begin{equation}\label{eq:spin_chain}
H=t\sum_j\sigma_j^z-t\sum_j\sigma_j^x \sigma_{j+1}^x-g\sum_j\sigma_j^z \sigma_{j+1}^z-g\sum_j\sigma_j^x \sigma_{j+2}^x.
\end{equation}

Despite the 
complicated form of the Hamiltonian~\eqref{HDir} and the absence of particle number conservation, we have solved this model for large system sizes 
using DMRG as well as by a combination of field theory/RG and mean field arguments.  We have determined the 
complete phase diagram for both signs of the interactions. Altogether we find 4 different stable phases, sketched in Fig.\ \ref{phaseD}(d). 
Without loss of generality, we may assume $t>0$. For $-0.285<g/t<250$, we find a critical ``Ising'' phase with a single gapless relativistic 
Majorana fermion excitation.  For $-2.86<g/t<-0.285$, we find a phase consisting of decoupled gapless Majorana and Luttinger liquid excitations, 
with central charge $c=3/2$ and a continuously varying Luttinger parameter $K<1$, corresponding to repulsive interactions.  
The Luttinger-liquid sector of this phase has a conserved charge, which is an emergent symmetry, not present in the lattice model. 
At strong coupling of either sign, gapped phases with spontaneously broken symmetries and degenerate ground states occur. 
For $g/t<-2.86$, the ground state is 4-fold degenerate. {The symmetry of the Hamiltonian~\eqref{eq:hamil} under translation by one Majorana site is spontaneously broken down to translation by 4 Majorana sites in the ground-state wave functions, giving rise to a unit cell of 4 sites. For $g/t>250$, on the other hand, the ground state is 2-fold degenerate with a unit cell of 2 sites. (Translation symmetry is broken down to translation by two Majorana sites). Since two Majoranas can be combined into one Dirac fermion as in Eq.~\eqref{HDir}, in terms of Dirac fermions, translation invariance is not broken in the gapped phase with positive $g$ ($g/t>250$). However, the Hamiltonian~\eqref{HDir} is symmetric under the particle-hole transformation $c_j\to c^\dagger_j$ at $t=0$. This additional particle-hole symmetry, only present for $t=0$, is spontaneously broken in this limit. }
The models with $g/t=\pm \infty$ are equivalent and we find that for large positive $g/t$ there is a low lying doublet of 
excited states, with energy per unit length $\propto |t|$, which become degenerate with the two ground states as $g/t\to \infty$.

There are 3 critical points in the phase diagram, as well as the first order transition at $g/t=\infty$. The transition from the Ising 
phase to the gapped 2-fold degenerate phase at $g/t=250$ is in the tricritical Ising universality class, with $c=7/10$. 
The corresponding CFT is supersymmetric. The relevant operator that drives the transition at this 
critical point respects the supersymmetry, which is therefore present in the gapped phase for $g/t$ slightly 
bigger than the critical value $g/t=250$.  (The effective Hamiltonian is also supersymmetric at  $g/t$ slightly smaller than this critical value 
but supersymmetry is spontaneously broken in this Ising phase.) On the negative-$g$ side, the transition between Ising and Ising plus Luttinger liquid (LL) 
phases at $g/t=-0.285$ is characterized by the vanishing of the velocities, giving rise to a dynamical exponent $z=3$ at the critical point, and an effective Fermi wave vector which is zero at the critical point and grows as we move into the Ising+LL phase. This critical point describes a Lifshitz transition as the number of low-energy Majorana modes changes from one (in the Ising phase) to three (in the Ising+LL phase), which is analogous to the change in the topology of the Fermi surface.

The dynamically induced Fermi wave vector continuously changes within the Ising+LL phase and reaches 
a commensurate value of $\pi /4$ at the third critical point, $g/t=-2.86$, where a transition occurs into 
a gapped 4-fold degenerate state. This is related to the commensurate-incommensurate (C-IC)
transition, which occurs in the spinless Dirac chain as the chemical potential is varied for $V>2t$~\cite{Haldane1980, Schulz1980}. 
It is, however, a nontrivial generalization of the C-IC transition since the interaction driving the transition 
couples LL and Ising sectors. We therefore term this transition generalized C-IC.

{Although the bulk of this paper deals with the translationally invariant (translation by one Majorana site) system described by the Hamiltonian~\eqref{eq:hamil}, a more general one-dimensional array of interacting Majoranas, in which we only have invariance under translation by two Majorana sites, may be relevant to experiments. This may occur, e.g., if the vortices are arranged in a dimerized pattern. In general, the phase diagram of the dimerized model will depend on three dimensionless parameters $t_2/t_1$, $g_1/t_1$, and $g_2/t_1$ instead of just one parameter $g/t$ [See Eq. (2.5)]. While an in-depth analysis of the phase diagram of the dimerized model is beyond the scope of this paper, we present a topological classification of the gapped phases of this model and  identify them using self-consistent mean-field calculation. 

We argue that according to the results of Refs.~\cite{Fidkowski2010,Fidkowski2011} the gapped phases of the dimerized model fall into eight categories characterized by the difference between the number of time-reversal even and odd Majorana bound states at the endpoints of the chain with open boundary conditions. Within the self-consistent mean-field scheme, we find that four of these topological classes can (and do in parts of the phase diagram) appear. The mean-field picture also provides a good description of certain features of the phase diagram of the nondimerized model. In particular, it accurately predicts the Lifshitz transition.

} 

The remainder of this paper is organized as follows. In the next section we discuss the weak and strong 
coupling limits (for both signs of the coupling constant, $g$). In Sec.\ III we analyze  the model with $g<0$ and 
in Sec.\ IV,  $g>0$.  In Sec.\ V we employ mean-field theory to  discuss a more general dimerized model, where $t$ and $g$ alternate, and present 
a topological classification of the gapped phases arising in the system.. Sec.\ VI elaborates on possible experimental realizations of our model and various signatures of its phases and phase transitions. Conclusions are given in Sec.\ VII and  further technical details appear in three Appendices.

\section{Weak and Strong Coupling Limits}
\subsection{Weak Coupling Limit}\label{sec:weak}

{Weak coupling treatment of interacting Majoranas relies on solving the noninteracting problem and treating the interactions with RG (see Refs.~\cite{Lai2011,Lobos2012,Cheng2014} for a few examples).}We first discuss the noninteracting case, $g=0$. As shown in Appendix ~\ref{ap:1}, the low-energy Hamiltonian of the noninteracting system is given by\bea \label{heff1}
& &H_0=iv\int dx [\gamma_L\partial_x\gamma_L-\gamma_R\partial_x\gamma_R],\ \  (v=4t)
\\ 
& &\{\gamma_{R/L}(x),\gamma_{R/L}(y)\}= (1/2)\delta (x-y),\eea
which describes free relativistic Majoranas. This is a CFT with central charge $c=1/2$, corresponding to the critical point of the transverse-field Ising model [see Eq.~\eqref{eq:spin_chain} with $g=0$].
Note that no mass term $H_m\equiv im\int dx \gamma_R\gamma_L$
appears. This is a consequence of the translational symmetry, $\gamma_j\to \gamma_{j+1}$ which maps
\be \gamma_R\to \gamma_R,\ \  \gamma_L\to -\gamma_L.\ee
Taking into account that $\gamma_{R/L}(x)$ vary slowly on the lattice scale, it is now straightforward to write the interaction term projected onto the low-energy states as
\be H_{int}\approx -256g \int dx \gamma_R\partial_x\gamma_R\gamma_L\partial_x\gamma_L.\ee
Note that 2 derivatives are necessary to get a nontrivial interaction by Fermi statistics. This 
interaction has RG scaling dimension 4,  1/2 for each Majorana field and 1 for each derivative,
and so is highly irrelevant.  [Dimension $\Delta >2$ is irrelevant in relativistic (1+1) dimensional field theory.]

This implies  the existence of an extended phase with a massless Majorana  in the vicinity of $g=0$.  Note 
that the present scenario is quite different from what happens in the 1D Hubbard and Dirac [see Eq.~\eqref{Dirac}] chains, where nonderivative operators can occur. 
These lead to the Mott transition at infinitesimal coupling in the Hubbard chain and to the continuously 
varying Luttinger parameter in the gapless phase of the spinless Dirac chain.

\subsection{Strong Coupling Limit}\label{sec:strong}
Unlike the spinless Dirac chain, the strong coupling ground states of the Majorana chain are nontrivial.  
However, insight into the nature of these ground states can be obtained by considering a model with 
alternating hopping and interaction terms:
\be
\begin{split} H=\sum_j&\left(it_1\gamma_{2j}\gamma_{2j+1}+it_2\gamma_{2j+1}\gamma_{2j+2}\right.\\
&+g_1\left.\gamma_{2j}\gamma_{2j+1}\gamma_{2j+2}\gamma_{2j+3}+
g_2\gamma_{2j+1}\gamma_{2j+2}\gamma_{2j+3}\gamma_{2j+4}
\right).\label{H'}
\end{split}\ee
We will argue that the symmetry of translation by one site is spontaneously broken at strong coupling and that the ground states are thus 
qualitatively similar to the ones for the Hamiltonian of Eq. (\ref{H'}).

\begin{figure}[t]
\centerline{\includegraphics[width=80mm]{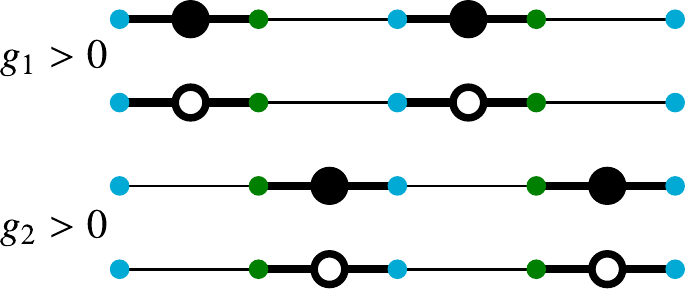}}
\caption{Mean-field ground states for $g>0$, corresponding to ferromagnetic states in the spin representation.  The small blue and green circles represent the sites of the Majorana chain. Bold links with large circles on them represent Dirac fermions formed by combining two Majoranas (two types of combinations are considered: blue-green and green-blue).  A filled circle corresponds to the  Dirac level being filled and an empty circle to it being empty.}
\label{g>}
\end{figure}

\begin{figure}[b]
\centerline{\includegraphics[width=80mm]{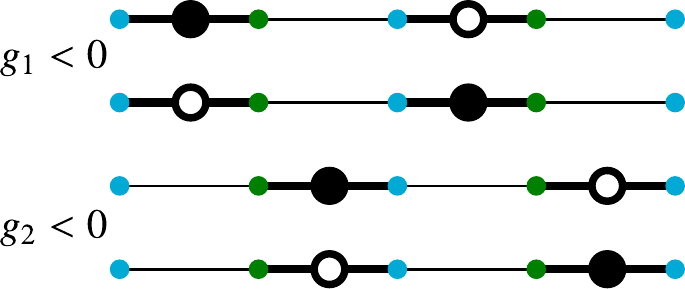}}
\caption{Mean-field ground states for $g<0$, corresponding to the antiferromagnetic states in the spin representation}
\label{g<}
\end{figure}

In the strong coupling limit $t_i=0$, the model is trivially soluble if either $g_1$ or $g_2=0$. For instance, when $g_2=0$
it follows from Eq.~\eqref{HDir} that
\be H\to -g_1\sum_j\hat p_j\hat p_{j+1}.\label{HSC}\ee
For $g_1>0$, corresponding to attractive interactions, the two ground states have all Dirac levels filled ($p_j=1$) or empty ($p_j=-1$). The case of $g_1>0$ corresponds to a ferromagnetic Ising interactions in the spin representation [see Eq.~\eqref{eq:spin_chain}]. For $g_1<0$, 
a charge density wave occurs, with every second Dirac level filled or empty, $p_j=\pm(-1)^j$ (antiferromagnetic spin chain). Similarly, if $g_1=0$, we combine Majoranas 
to form Dirac operators on sites $(2j+1,2j+2)$ obtaining the same Eq. (\ref{HSC}) for a different set of Dirac fermions (shifted by one Majorana site).  These 4 different ground states are indicated 
in Figs.\ \ref{g>} and \ref{g<} for the cases $g_i>0$ and $g_i<0$ respectively.  Clearly, there is a  gap, $2|g_i|$,  
to the lowest energy excited states. 
While we can only solve the model exactly 
at strong coupling when the interactions vanish on half the quartets of Majoranas, we will see that the 
4 corresponding ground states, for either sign of $g_i$, are cartoon representations of the actual ground states 
of the uniform chain in the strong coupling limit. If this is correct, there are two spontaneously broken discrete symmetries. 
One of these is translation by 1 site: $\gamma_j\to \gamma_{j+1}$. 
In the special case $t=0$, the model has particle hole symmetry, $c_j\to c_j^\dagger$.  For both definitions of 
Dirac modes this corresponds, up to a phase, to
\be \gamma_j\to (-1)^j\gamma_j.\ee
Clearly this particle-hole symmetry is spontaneously broken when $t_1=t_2=0$ and either $g_1$ or $g_2=0$.  The two ground states in 
those cases are mapped into each other by a particle-hole transformation.  We expect that it is 
also spontaneously broken when $g_1=g_2$ and $t_1=t_2=0$.  The combination of spontaneously broken translation symmetry and 
particle-hole symmetry results in 4-fold ground state degeneracy. 

It is also important to note that, for $t=0$, the duality transformation:
\bea \gamma_j&\to& -\gamma_j,\ \  (j=4n)\nonumber \\
&\to& \gamma_j,\ \  (\hbox{otherwise})\eea
maps the Hamiltonian $H\to -H$ corresponding to $g\to -g$.  This transformation interchanges the cartoon ground states 
of Figs. (\ref{g>}) and (\ref{g<}). 
Thus if the spontaneously broken ground states occur for one sign of $g$ they must occur for the other.

{We note that the Hamiltonian~\eqref{H'} can be exactly mapped into a spin Hamiltonian by a Jordan-Wigner transformation, giving
 \begin{equation}\label{eq:spin_chainD}
H=t_1\sum_j\sigma_j^z-t_2\sum_j\sigma_j^x \sigma_{j+1}^x-g_1\sum_j\sigma_j^z \sigma_{j+1}^z-g_2\sum_j\sigma_j^x \sigma_{j+2}^x.
\end{equation}
The special case $g_1=0$ corresponds to the anisotropic next-nearest-neighbor Ising (ANNNI) model. 
(For a review see~\cite{Selke1988}.) For $t_1=t_2=0$, the Hamiltonian \eqref{eq:spin_chainD} maps to a multispin Ising chain with $H=-g_1\sum_j\tau_j^z-g_2\sum_j\tau_{j-1}^x\tau_{j}^x\tau_{j+1}^x\tau_{j+2}^x$ through a transformation $\tau_j^z=\sigma_j^z\sigma_{j+1}^z$. This model has been studied in the literature \cite{Turban1982,Penson1982,Alcaraz1986,Blote1986} and based on a connection with the 8-state Potts model as well as some numerical evidence \cite{Alcaraz1986,Blote1986} is believed to have a first-order transition at the self-dual point $g_1=g_2$. Therefore, (i) the system is likely gapped for $t_1=t_2=0$, $g_1=g_2$ and (ii) the ground states on two sides of the transition (respectively dominated by $g_1$ and $g_2$ only) are present at the self-dual point. In the proceeding sections, we provide numerical evidence for this scenario.}


Assuming these gapped broken symmetry ground states at $t=0$, let us now consider the likely effect of a small 
nonzero $t$. It is again convenient to start with the case $g_2=0$, $g_1=g$ and turn on only a small $t_1$ in 
the dimerized Hamiltonian of Eq. (\ref{H'}). Again, in Dirac notation, we have a special case of Hamiltonian~\eqref{HDir}:
\be\label{Hdir2} H=\sum_j\left(t_1\hat p_j-g_1\hat p_j\hat p_{j+1}\right).\ee
We see that $t$ breaks particle-hole symmetry, favoring empty Dirac levels for $t_1>0$ or 
 filled for $_1<0$.  Its effects are quite different depending on the sign of $g_1$. For $g_1>0$, $t_1$ splits the degeneracy between the two  ground states, favoring the state with all Dirac levels empty or filled, depending on its sign.  This would correspond to a first-order phase transition with a jump in $\langle \hat p_j\rangle$ (and consequently a jump in the occupation number $\langle \hat n_j\rangle$ as $\hat p_j=2\hat n_j-1$) at $t_1=0$ {(see Appendix C for numerical evidence).}
 On the other hand, for $g_1<0$,  turning on a small $t_1$ does not split the degeneracy of the two 
  ground states which remain the exact ground states for $|t_1|<|g_1|$, at which point a {transition 
occurs (through a multicritical point~\footnote{For $\pm t_1 = g_1<0$, there is zero energy cost to add a soliton between the
two charge-density-wave states since an energy $-|t_1|$ is gained by flipping
the occupancy of a site and an energy $|g_1|$ is lost from having 2
neighboring sites that have the same occupancy. Thus any state with an
arbitrary configuration of solitons and antisolitons has the same energy.
}) into one }of the $g_1>0$ ground states with all Dirac levels empty or filled, depending on the sign of $t_1$. The Ising spin-chain representation $\hat p_j=\sigma^z_j$ is also illuminating: the term proportional to $t_1$ in Eq.~\eqref{Hdir2} serves as a magnetic field coupled to the magnetization, which splits (does not split) the degeneracy of a ferromagnet (antiferromagnet) when it is less than the  exchange 
coupling. Note that, in this sub-section, we have {\it explicitly} broken the symmetry of translation by one Majorana lattice site. 
Given this assumption, we have argued that, for $g<0$, but not $g>0$, there is a further {\it spontaneous} breaking 
of symmetry of translation by two Majorana sites.
We will argue that these results are qualitatively correct for the translationally invariant Hamiltonian, with $t_1=t_2$, $g_1=g_2$, 
In the strong coupling regime, $|g|\gg |t|$, there is a 2-fold degenerate ground state for large positive $g$ and a 
4-fold degenerate ground state for large negative $g$. 
However, we shall see that the transition out of the  strong coupling phases is continuous, for both signs of $g$.

\section{Repulsive Interactions: $g<0$}\label{sec:repul}
\begin{figure}
\centerline{\includegraphics[width=80mm]{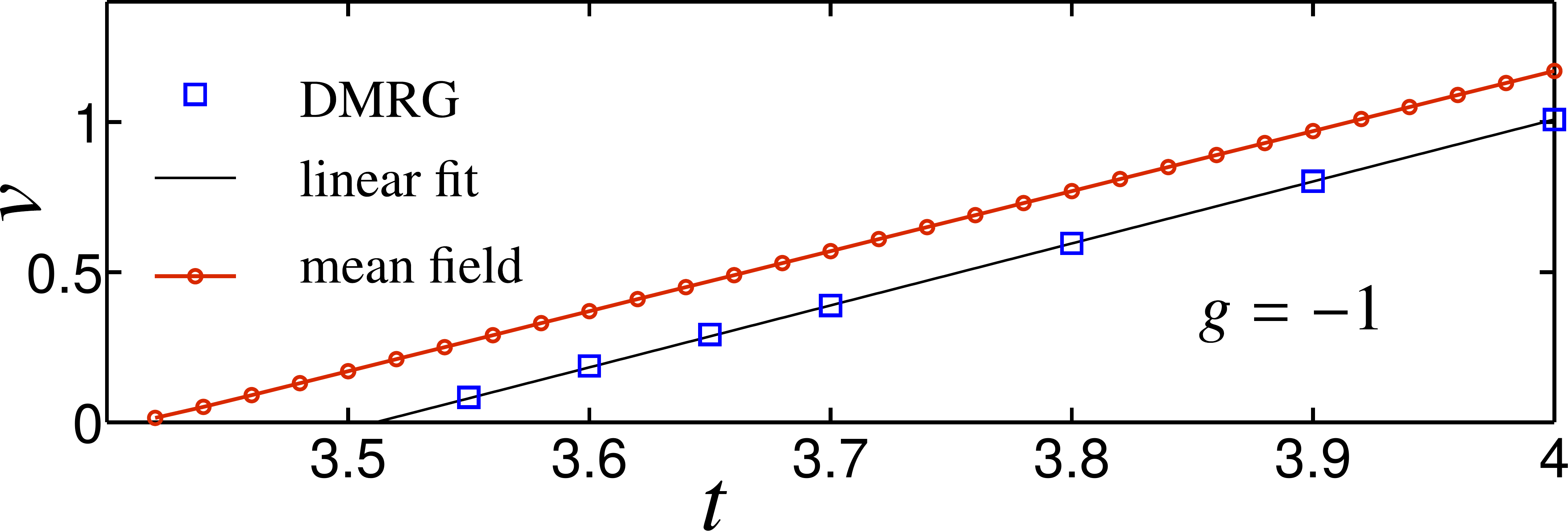}}
\caption{Velocity in the Ising phase near the Lifshitz transition. As the transition is driven by a renormalization of the dispersion relation, mean-field calculations are in approximate agreement with DMRG.}
\label{vg<}
\end{figure}

As discussed in Sec.~\ref{sec:weak}, we expect the Ising phase to persist up to a finite critical $g/t$ of either sign.  We have 
confirmed this primarily by calculating several energy levels in the finite-size spectrum, with antiperiodic boundary conditions (APBC). 
We can classify all states by their fermion parity which we formally define, for a chain with an even number of Majorana sites $L=2\ell$ numbered $0,1,2, \ldots (L-1)$, as
\be F=\prod_{j=0}^{\ell-1}\left(i\gamma_{2j}\gamma_{2j+1}\right).\ee
Note that, defining Dirac fermions by Eq. (\ref{cdef}), this becomes $F=\prod_{i=0}^{\ell-1}\hat p_j=(-1)^{N_F+\ell}$,
where $N_F$ is the number of Dirac fermions added to the vacuum state. [While $N_F$ is not conserved, $(-1)^{N_F}$ is in Hamiltonian~\eqref{HDir}.]
In the Ising, i.e., free Majorana, phase, the low-energy excitations with APBC simply correspond to creating particles of energy 
$Ev|k|$, with $k={2\pi v}(n+1/2)/L$
for integer $n$.  Thus the lowest energy excited state has opposite fermion parity to the ground state,  
energy $\pi v/L$ and is two-fold degenerate. The lowest excited state with the same fermion parity as the ground state has 
2 fermions added at $k=\pi /L$ and $k=\pi-\pi /L$ with energy $2\pi v/L$ ($k$ is restricted to the interval $0\leq k<\pi$. See Appendix A.).  This ratio of excitation energies, two, remains essentially constant 
in our DMRG results over the entire Ising phase while the velocity, $v$ varies.  We performed the computations at $g=1$ by varying $t$. As $t/g\to -3.512$ (corresponding to $g/t=-0.285$), we find that the velocity 
vanishes as shown in Fig.\ \ref{vg<}.  As we decrease $t$, the behavior of the finite-size spectrum changes (compared with the Ising phase realized at larger $t/g$). As we see this is a Lifshitz transition corresponding to a change in the topology of the Fermi surface. {The velocities were extracted from a linear fit of the finite-size gaps computed with DMRG to $1/L$ for a range of system sizes with $L=40,\dots 200$ Majoranas. We kept 500 states in the DMRG computations to achieve convergence in the energy gaps.}

\begin{figure}
\centerline{\includegraphics[width=80mm]{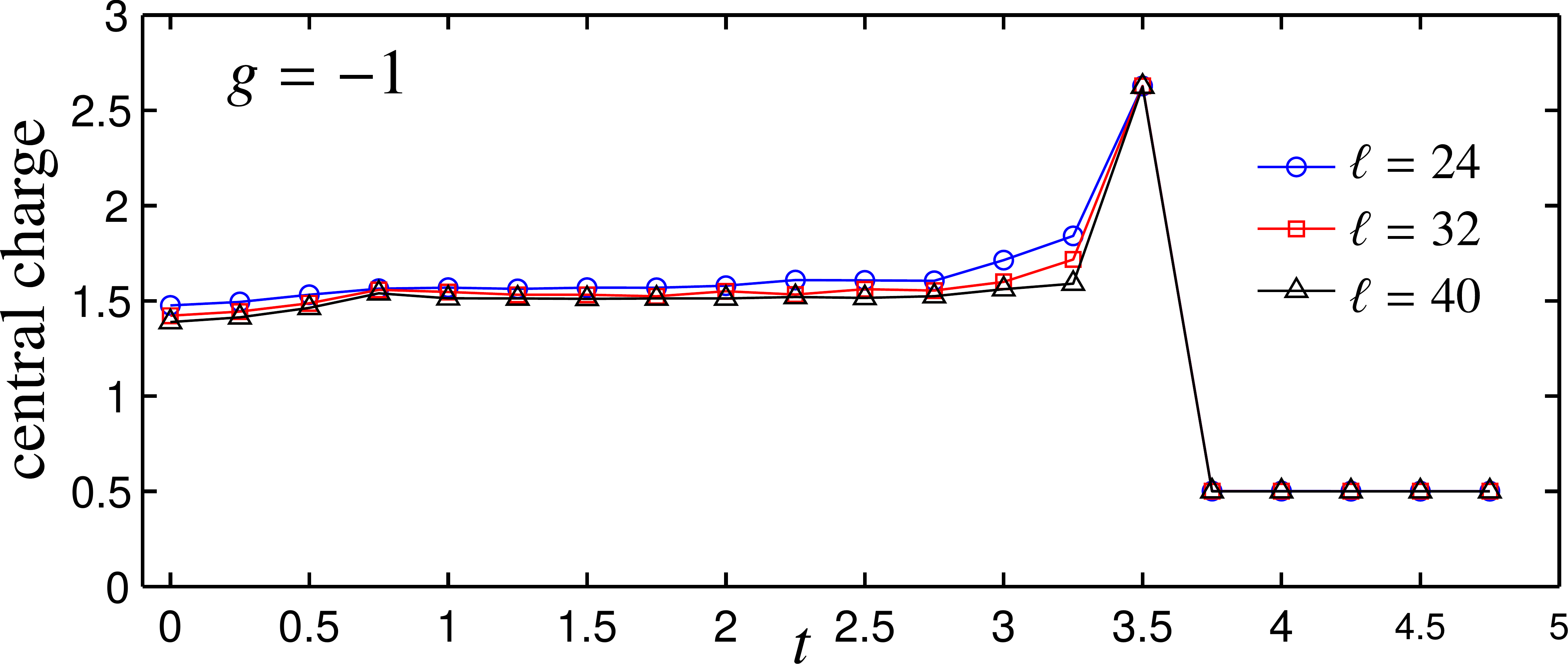}}
\caption{The behavior of the central charge as a function of $t$ for $g=-1$.\label{fig:central}}
\end{figure}

Computing the central charge sheds light on the nature of the phase for larger hopping $t$. {In a critical phase described by a $(1+1)d$ CFT,
the entanglement entropy $S$ of a subsystem of length $y$ with the rest of the system is related to the central charge $c$ through
\be S={c\over 3}\log\left[{L\over \pi a}\sin\left({\pi y\over L}\right)\right]+{\rm const.}\ee
 for a system of length $L$ with periodic boundary conditions~\cite{Calabrese2004}. We numerically computed the entanglement entropy with DMRG (keeping 300 states) and extracted the central charge from the relationship above.} As expected, the central charge in the Ising phase is $c=1/2$. However, as seen in Fig.~\ref{fig:central}, it jumps to $c=3/2$ at the same value of $t$, for which the velocity goes to zero. This indicates that {upon decreasing the value of $t$ beyond this transition point, }three species of low-energy Majoranas appear at this phase transition. In fact, as discussed below, the same behavior can arise in a noninteracting model with third-neighbor hopping. Therefore, we can understand this phase transition simply in terms of a kinetic energy renormalized by interactions. Incidentally, as we will see in Sec.~\ref{sec:dimer}, a mean-field calculation captures this transition with good accuracy (see Fig.~\ref{vg<}).

\begin{figure}
\centerline{\includegraphics[width=80mm]{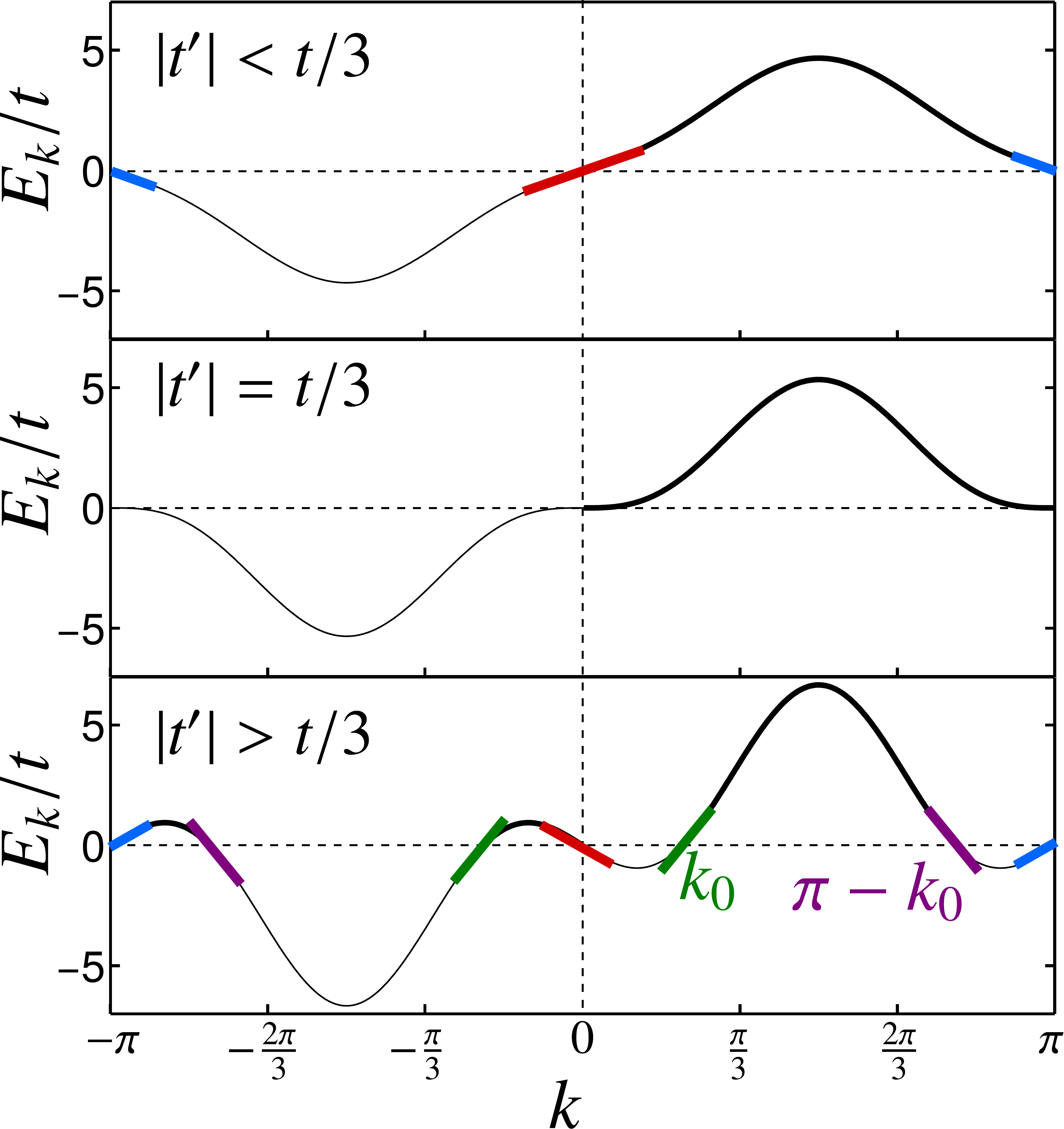}}
\caption{Dispersion relation indicating a Lifshitz transition. For $|t'|<t/3$, the zeros of the dispersion relation are at $k=0$ and $k=\pi$. At $|t'|=t/3$ the velocities of both these low-energy modes vanish and for $|t'|>t/3$, new low-energy modes appear at finite momenta $k_0$ and $\pi-k_0$, with $k_0=0$ at the Lifshitz transition $|t'|=t/3$.\label{MFdisp}}
\end{figure}

Third-neighbor hopping is indeed allowed by all symmetries. [Spatial parity symmetry $\gamma_j\to (-1)^j\gamma_{-j}$
forbids a second neighbor hopping term. {Notice that a naive parity transformation $\gamma_j\to \gamma_{-j}$ changes the sign of the nearest-neighbor hopping term and the $(-1)^j$ term simply correct for this.]} Consider a quadratic Hamiltonian
\be H=i\sum_j\gamma_j[t\gamma_{j+1}+t'\gamma_{j+3}]={1\over 2}\sum_k E_k\gamma (-k)\gamma (k)\label{eq:3rd}\ee
with $E_k=4t\sin k+4t'\sin (3k)$.
As in Appendix ~\ref{ap:1}, it is convenient to regard $\gamma (k)$ as an annihilation operator for the regions of $k$ where $E_k>0$ and write 
$\gamma (k)$ as $\gamma^\dagger (-k)$ for the complementary regions.  Consider the case $t>0$, $ t'<0$. For $t'>-t/3$, $E_k$  vanishes 
at $k=0$ and $\pi$ only, with velocity $v=4t+12t'$. However, $v\to 0$ at $t'=-t/3$.  For $t'<-t/3$, $E_k$ vanishes at 4 other points, 
$\pm k_0$ and $\pm (\pi -k_0)$ with $\sin k_0=(1/2)\sqrt{3+t/t'}$.
Now there are 3 regions of $k$ where $E_k>0$, shown by thick black lines in Fig. (\ref{MFdisp}). 
The velocity at $k=0$ is $v_0=16\sin^2k_0$, while at $k=k_0, \pi-k_0$, we have $v=2v_0\cos k_0$.
Note that $v_0$ and $v$ increase linearly with $-t'-t/3$ while $k_0$ increase more rapidly $\propto \sqrt{-t'-t/3}$.  
Here, $k_0$ plays the role of a Fermi wave vector.
 We may again introduce relativistic fermions to represent the low-energy excitations as
\be \gamma_j \approx 2\gamma_L(j)+(-1)^j2\gamma_R(j)+\left[e^{-ik_0j}\psi_R(j)+e^{i(k_0-\pi )j}\psi_L(j)+{\rm H.c.}\right].
\label{glow}\ee

Here $\psi_{R/L}$ are Dirac fermion operators, {simply related to the Fourier modes of the original Majoranas as
\begin{equation}\label{eq:RL}
\psi_R(q)=\gamma(k_0+q),\quad \psi_L(-q)=\gamma(\pi-k_0-q), \quad -\Lambda<q<\Lambda,
\end{equation}
where $\Lambda\ll 1$ is the  momentum cut-off of the low-energy sector. }Note that the right/left movers occur at $k$ points where $E_k$ has positive/negative slope. 
For $k$ slightly larger than $k_0$, $\gamma (k)$ is identified with a right-moving particle annihilation operator whereas for $k$ slightly 
less than $k_0$ it is identified with a right-moving antiparticle creation operator.  The low-energy Hamiltonian becomes
\be H_0=i\int dx [v_0(\gamma_L\partial_x\gamma_L-\gamma_R\partial_x\gamma_R)+v(\psi^\dagger_L\partial_x\psi_L-\psi^\dagger_R\partial_x\psi_R)].\ee

We now consider the effect of the interactions. These are most rigorously treated if we added a $t'$ term to the Hamiltonian by hand, and then 
turned on a small $g$. However, we expect the universal properties of the resulting phase to also describe the case at hand where $t'$ 
is generated dynamically. But, in this case we are not in the weak coupling regime since $|g|$ must be ${\cal O}(t)$ to drive the 
Lifshitz transition.  Since we have more fields in the low-energy field theory, it is possible to have nonderivative interaction terms. 
Many of these come with spatially oscillating factors, making them irrelevant for general values of $k_0$. However, there are 
two nonoscillatory 4-fermion interactions allowed by symmetry:
\be H_{int}\approx \int dx\left[g_0 :\psi^\dagger_L\psi_L\psi^\dagger_R\psi_R:+{ g'} \gamma_R\gamma_L\left(\psi_L\psi_R+\psi_L^\dagger \psi_R^\dagger \right)
\right],\label{Hint}\ee
where $g_0=-16g[\cos k_0-\cos (3k_0)]$ for weak coupling and ``$:$'' indicates normal ordering. Since we are considering $g<0$, we have  $g_0>0$ corresponding to repulsive interactions.  
The effects of this term by itself are well-known and easily treated using bosonization techniques, leading to a Luttinger liquid (LL). This 
corresponds to a free massless relativistic boson theory with the RG scaling dimensions varying continuously.  These scaling dimensions 
are controlled by a single dimensionless parameter $K$ known as the Luttinger parameter, which takes the value $K=1-{g_0\over 2\pi v}+\ldots$ for weak coupling. Generally, we have $K<1$ for repulsive interactions.

We now argue that the second term in Eq.~\eqref{Hint} above is irrelevant in the RG sense for $K<1$. The scaling dimension of $\psi_L\psi_R$ appearing in the $g'$ interaction is $1/K$, which is larger than one for repulsive interactions. 
The $\gamma_R\gamma_L$ factor in this term also contributes 1 to the scaling dimension which leads to $\Delta =1+1/K>2$, making it irrelevant.  Therefore, we find that the Ising and LL sectors 
are decoupled in the low-energy theory.  This implies that a $U(1)$ charge conservation symmetry emerges in the LL sector of the Ising+LL phase, along with an associated Fermi wave-vector $k_0$. The charge is related to the occupation of modes near momenta $\pm k_0$ and $\pm (\pi-k_0)$ as 
\begin{equation}\label{eq:U(1)}
\hat{N}=\sum_{-\Lambda<q<\Lambda}\left[\gamma^\dagger(k_0+q)\gamma(k_0+q)+\gamma^\dagger(\pi-k_0-q)\gamma(\pi-k_0-q)-1\right].
\end{equation}
{The constant $-1$ in the sum above is chosen such that there are no fermions in the ground state. For example, instead of writing the number of right movers as $\sum_ {-\Lambda<q<\Lambda}\psi_R^\dagger(q)\psi_R(q)$, which is nonzero in the ground state (with $q<0$ modes occupied), we write 
\begin{equation}
\hat{N}_R=\sum_{0<q<\Lambda}\psi_R^\dagger(q)\psi_R(q)-\sum_{-\Lambda<q<0}\psi_R(q)\psi_R^\dagger(q),
\end{equation}
which measures the charge of the right-movers with respect to the Fermi level. The above expression and its analog for the left-movers leads to Eq.~\eqref{eq:U(1)} through $\hat{N}=\hat{N}_R+\hat{N}_L$ [see Eq.~\eqref{eq:RL}].}

All of this is in good agreement with our DMRG results. Again, the finite-size spectrum provides a powerful technique 
for confirming the phase diagram and extracting the universal parameters, which govern it. As shown in Appendix \ref{ap:2}, the excitation spectrum of the system for a noninteracting LL ($K=1$) can be computed using elementary methods and is given by
\be \Delta E={2\pi \over L}\left[ {v\over 4}(N-k_0L/\pi )^2+{v\over 4}M^2
+ {v_0\over 4}N_I^2+{v_0\over 4}M_I^2\right],
\label{fss}\ee
where $N$ and $M$ ($N_I$ and $M_I$) are \textit{integers of the same parity}, labeling  excitations  in the LL (Ising) sector. Physically, $N$ is the total number of particles (both right and left movers), while $M$ is the difference between the number of right and left movers, characterizing the charge current (this also implies that $N$ and $M$ have the same parity). {In the ground state, we either have $M=0$ and $N$ is the closest even integer to $k_0L/\pi $ or $M=\pm 1$ with $N$ is the closest odd integer. The value of $N$ in the ground state determines how much it is shifted from the eigenvalues of the operator $\hat N$ in Eq.~\eqref{eq:U(1)}.} We have neglected the particle-hole 
excitations, which do not contribute to the low-energy excited states we study in this paper.

The effect  of the marginal coupling constant $g_0$ in Eq. (\ref{Hint}) on the finite-size spectrum can be accounted for as follows. 
After bosonizing, the charge density becomes proportional to the spatial derivative of a massless boson field, $\partial_x\phi$. Similarly, the current density becomes proportional to the derivative of its dual boson, $\partial_x\theta$. The Luttinger parameter 
scales $\phi$ by $1/\sqrt{K}$ and $\theta$ by $\sqrt{K}$. It then follows that Eq. (\ref{fss}) becomes 
\be \Delta E={2\pi \over L}\left[ {v\over 4K}(N-k_0L/\pi )^2+{vK\over 4}M^2
+ {v_0\over 4}N_I^2+{v_0\over 4}M_I^2\right].
\label{fssI}\ee
There is no change in the energy of Ising excitations due to the interactions, since they only act in the LL sector.

\begin{figure}
\centerline{\includegraphics[width=80mm]{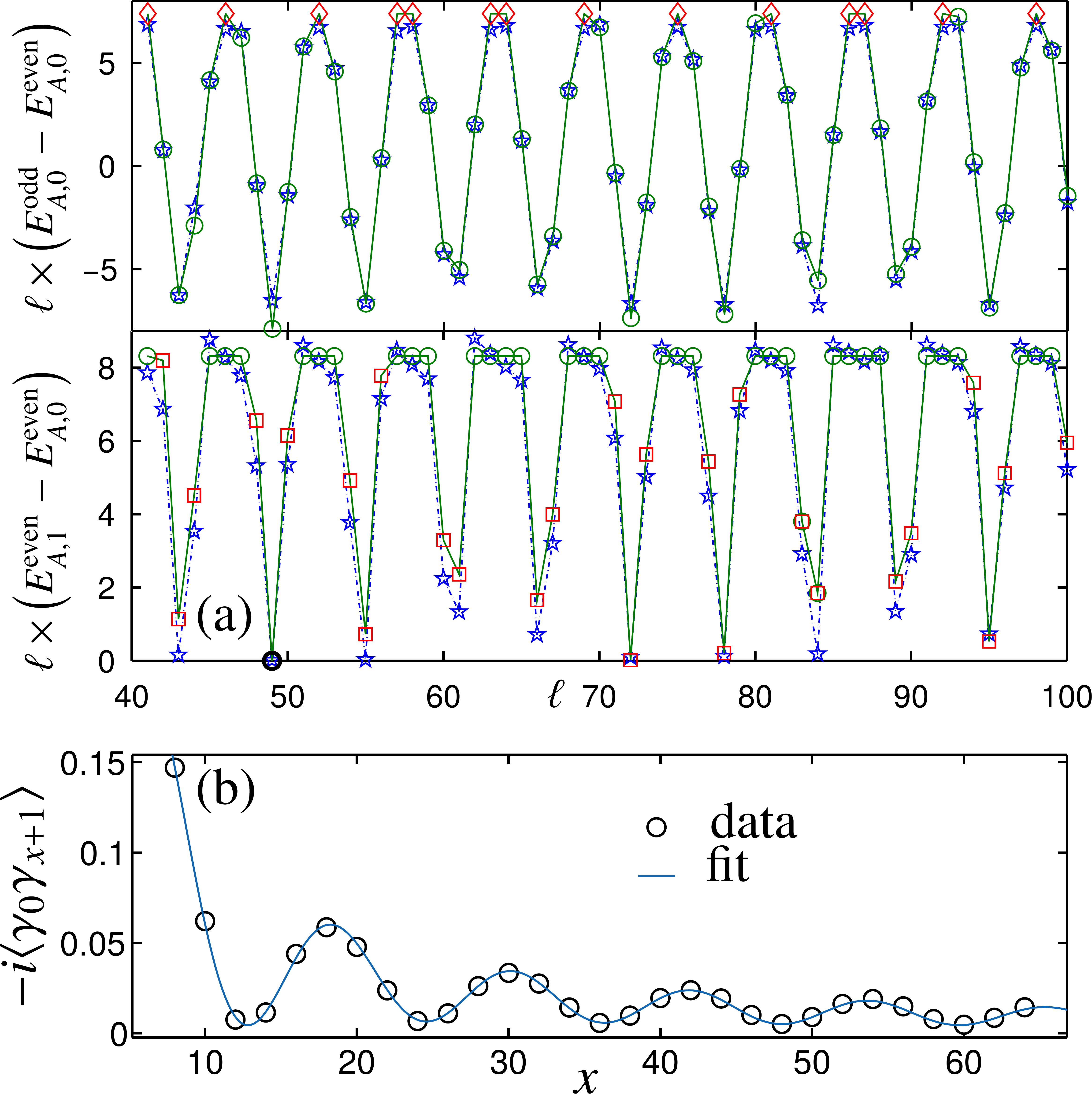}}
\caption{(a) Odd-even and even-even gap, scaled by $L$ for $g=-1$ and $t=2.25$.  The stars are DMRG data and the other symbols are 
the best fit to the LL+Ising spectrum. The blue symbols correspond to excitations in the LL sector whereas the red ones 
correspond to excitations in the Ising sector.  The black circle corresponds to an excited state of even parity containing 
simultaneous Ising and LL excitations. (b) The Majorana correlation function. The fit to spectrum give $K=0.4517$ and $k_0=0.5444$, while fitting the correlation function (for even $x$ so that there are no $(-1)^x$ oscillations) gives $K=0.4611$ and $k_0=0.5375$, in very good agreement. {The DMRG results were checked for convergence (with negligible truncation error) for each system size. The maximum number of states kept in the computations was 700.}}
\label{fit}
\end{figure}

The finite-size spectrum now exhibits great complexity as $k_0L/\pi$ is varied, either by varying $L$ or $k_0$ via the hopping parameter $t$ 
(with $g$ held fixed at $g=-1$). {We are interested in the ground state and the first excited state in the two sectors corresponding to the total fermion parity. While the absolute ground state must have $N_I=M_I=0$, we may have Ising excitations in the ground state in a given parity sector. To proceed, we first define $f_{\rm even/odd}(x)$ as the closest even/odd integer to a real number $x$.  It is easy to show that these functions are given by
\begin{eqnarray}
f_{\rm even}(x)&=&{1\over 2}\left(2\lfloor x\rfloor+1-(-1)^{\lfloor x\rfloor}\right),\\
f_{\rm odd}(x)&=&{1\over 2}\left(2\lfloor x\rfloor+1+(-1)^{\lfloor x\rfloor}\right),
\end{eqnarray}
where $\lfloor x\rfloor$ is the floor function of $x$.

Now, fixing the fermion parity, we can write the candidates for the ground state energy in each sector as
\begin{eqnarray}
E_{0, {\rm Dirac}}^{\rm even}&=&{2\pi\over KL}\left[f_{\rm even}({k_0\over \pi}L)-{k_0\over \pi}L\right]^2,\\
E_{0, {\rm Ising}}^{\rm even}&=&{2\pi\over KL}\left[f_{\rm odd}({k_0\over \pi}L)-{k_0\over \pi}L\right]^2+{\pi Kv\over 2L}+{\pi v_0\over L},\\
E_{0, {\rm Dirac}}^{\rm odd}&=&{2\pi\over KL}\left[f_{\rm odd}({k_0\over \pi}L)-{k_0\over \pi}L\right]^2+{\pi Kv\over 2L},\\
E_{0, {\rm Ising}}^{\rm odd}&=&{2\pi\over KL}\left[f_{\rm even}({k_0\over \pi}L)-{k_0\over \pi}L\right]^2+{\pi v_0\over L},
\end{eqnarray}
where the superscripts indicate the total fermion parity and the subscript Dirac (Ising) denoted the absence (presence) of an Ising excitation in the ground state of the given sector. We then have
\begin{eqnarray}
E_{0}^{\rm even}&=&\min(E_{0, {\rm Dirac}}^{\rm even},E_{0, {\rm Ising}}^{\rm even}),\\
E_{0}^{\rm odd}&=&\min(E_{0, {\rm Dirac}}^{\rm odd},E_{0, {\rm Ising}}^{\rm odd}).
\end{eqnarray}

The absolute ground state energy is given by $\min(E_{0}^{\rm even},E_{0}^{\rm odd})=\min(E_{0, {\rm Dirac}}^{\rm even},E_{0, {\rm Dirac}}^{\rm odd})$. When $k_0L/\pi$ is close to an even integer, 
the ground state has even fermion parity with $N$ even and $M=0$. However, if $k_0L/\pi$ is close to an odd  
integer, the two-fold degenerate ground states have odd fermion parity with $N$ an odd integer and $M=\pm 1$. 
As $K$ gets smaller, the latter scenario occurs for a larger range of $k_0L/\pi$ determined from the condition
\begin{eqnarray}
K^2<{8\over v}(-1)^{\lfloor k_0L/\pi \rfloor}\left(k_0L/\pi-\lfloor k_0L/\pi \rfloor-1/2\right).
\end{eqnarray}

 The lowest excited state in the odd 
sector exhibits even greater complexity. For some ranges of parameters it can be in the Ising sector, with $N_I=M_I=1$ 
and for other ranges in the LL sector. This can be determined by comparing the energies of several candidates as in the ground-state case. A sample of such data, for $t=2.25$ is shown in Fig.\ \ref{fit}. We can fit the data using the 4 parameters 
$v_0$, $v$, $k_0$ and $K$. As expected, the fit gets better as $L$ increases. 

}

We also verified that the fermionic correlation functions are in agreement with the spectrum. We expect from Eq.~\eqref{glow} that
\be\label{eq:corr}
i\langle\gamma_0\gamma_{x+1}\rangle\sim a/x+b\sin\left(k_0x+\phi\right)/x^{\left(K+1/K\right)/2}
\ee
for even $x$, nonuniversal coefficients $a$ and $b$, and phase shift $\phi$. Using even $x$ simplifies the fit as further oscillations with wave vector $\pi$ and $\pi-k_0$, i.e., $(-1)^x/x$ and $(-1)^x\sin\left(k_0x+\phi'\right)/x^{\left(K+1/K\right)/2}$ [see Eq.~\eqref{glow}], do not appear for even $x$. The values of $K$ and $k_0$ extracted from this are in agreement with those obtained from the finite-size spectrum.

\begin{figure}
\centerline{\includegraphics[width=80mm]{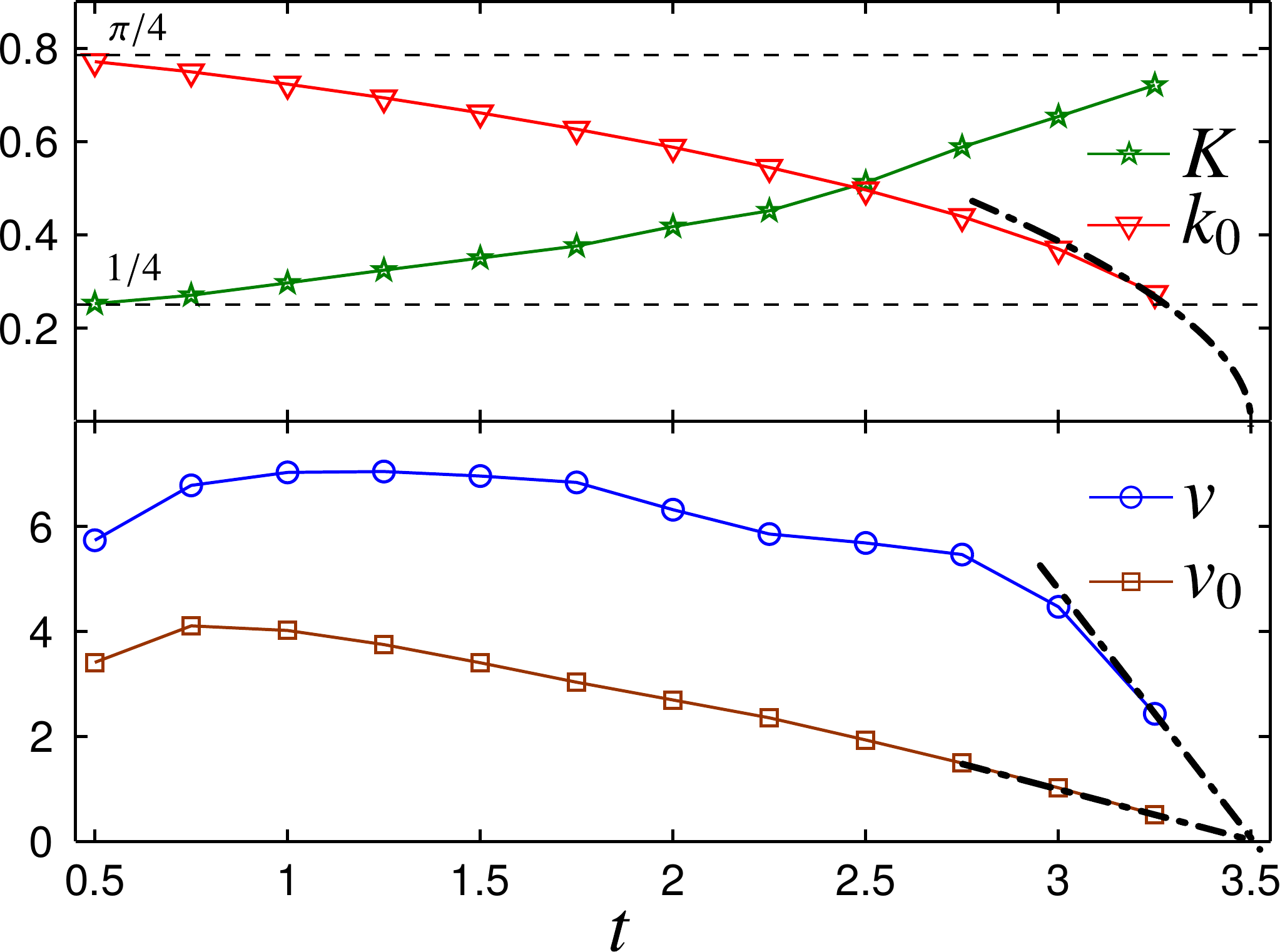}}
\caption{Parameters of LL+Ising phase determined from fitting the even-even and even-odd gaps, for a range of $t$, with $g=-1$.}
\label{param}
\end{figure}

We extracted the 4 fitting parameters from the finite-size spectrum and the results are plotted versus $t$ in Fig.\ \ref{param}. Note that 
both $v$ and $v_0$ appear to increase linearly with $t_L-t$ while $k_0$ increases as $\sqrt{t_L-t}$ as expected 
from the noninteracting model. $t_L$ is the value of $t$ at the Lifshitz transition for $g=-1$. 
Also $K$ decreases with increasing $|g|/t$ as expected for increasing repulsive interaction strength.

As we decrease $t$ to around $0.5$, the quality of fits decreases. We expect a phase transition to a gapped phase for some $t_C<0.5$. Importantly, it appears that $k_0\to \pi /4$ and $K\to 1/4$ as we approach the phase transition and fitting to the Ising+LL begins to fail. 
This transition can be understood by observing that a possible interaction term, which was not included in Eq.~\eqref{Hint} due to its fast oscillations at generic $k_0$, namely
\be H'\propto \int dx \gamma_R\gamma_L\left[e^{i(4k_0-\pi)x}\psi^\dagger_R\partial_x\psi^\dagger_R\psi_L\partial_x\psi_L-{\rm H.c.}\right],\ee
becomes nonoscillatory at $k_0=\pi/4$. 
The scaling dimension of $H'$ is $1+4K$. 
Our DMRG data in Fig.\ \ref{param} indicates that $K\to 1/4$ at the transition point so this interaction becomes relevant.  Since $H'$ couples together 
LL+Ising fermions, we expect to gap out both sectors.  The fact that its dimension goes to 2 at precisely the point where 
it becomes commensurate appears to be a novel generalization of the commensurate-incommensurate transition, 
which occurs in the spinless Dirac chain~\cite{Haldane1980,Schulz1980}.

To determine the critical value of $t_C$ for the C-IC transition, we utilize an exact degeneracy of the gapped phase with periodic boundary conditions (PBC) as discussed below. For smaller $t$, we expect that the qualitative description in Sec.~\ref{sec:strong} applies, with a 4-fold degenerate gapped ground state. 
The gap appears to remain very small all the way to $t=0$ making this phase challenging to study with DMRG. However, 
we found that the 4-fold ground state degeneracy provides a convenient way of determining $t_{C}$ accurately.  
This degeneracy turns out to be exact for $L=8N$ sites with PBC, where
all states come in degenerate pairs of opposite fermion parity.
This follows, for periodic boundary conditions (PBC), since translation by  
one site, $T$, maps $F$ into $-F$:
\be \label{eq:FF}F=\gamma_0\gamma_1\ldots \gamma_{2L-1}\to \gamma_1\gamma_2\ldots \gamma_{2L-1}\gamma_0=-F.
\ee
It then follows that for any energy eigenstate $|\psi\rangle$, $T|\psi \rangle$ is also an eigenstate of the same energy and 
opposite fermion parity. The full 4-fold degeneracy of the ground state is less generic and signals the broken symmetry phase. 
It corresponds to 2-fold degeneracy of the ground states in both even and odd fermion parity sectors. Let us focus, 
for example, on the even fermion parity sector. It is then convenient to start with the case $g_2=t=0$, so that 
the Hamiltonian only contains the $g_1$ term. Then the two ground states are known exactly:
\bea |\psi_1\rangle&=&c^\dagger_1c^\dagger_3\ldots c^\dagger _{4N-1}|0\rangle \nonumber \\
|\psi_2\rangle&=&c^\dagger_2c^\dagger_4\ldots c^\dagger _{4N}|0\rangle .
\eea
Now consider translation by 2 sites (in the Majorana chain), a symmetry even when only $g_1\neq 0$:
\bea T^2|\psi_1\rangle &=&|\psi_2\rangle \nonumber \\
T^2|\psi_2\rangle &=&c^\dagger_3c^\dagger_5\ldots c^\dagger _{4N-1}c^\dagger_1|0\rangle=-|\psi_1\rangle .
\eea
Thus we can construct linear combinations of these degenerate ground states:
\be |\psi_\pm \rangle \equiv (|\psi_1\rangle\mp i |\psi_2\rangle)/\sqrt{2}.\ee
which are eigenstates of $T^2$ with eigenvalues $\pm i$. This corresponds to momentum $\pm \pi /2$ with 
respect to translation by 2 Majorana sites, corresponding to 1 Dirac site.  That is, 
we define momentum $P'$ in this case by $T^2\equiv e^{iP'}$. 
Now consider gradually increasing $g_2$ to $g_1$.  Although the 2 ground states become more complicated they 
must remain degenerate since they have equal and opposite momentum, by spatial parity symmetry.  
At $g_2=g_1$ 
the degeneracy becomes 4-fold due to the additional symmetry of translation by one site.
Now defining 
momentum by $T=e^{iP}$, we can form 
linear superpositions of even and odd fermion parity ground states with momentum $\pm \pi /4$, $\pm 3\pi /4$. 
 Turning on a small $t$, the ground state degeneracy must survive up to 
the phase transition at $t_C$ by the same argument.  On the other hand, if the number of Majorana sites is $L=8N+4$ 
this argument fails, and mixing and splitting of the 2 low-lying states in each fermion parity sector occurs for finite system size.

\begin{figure}
\centerline{\includegraphics[width=80mm]{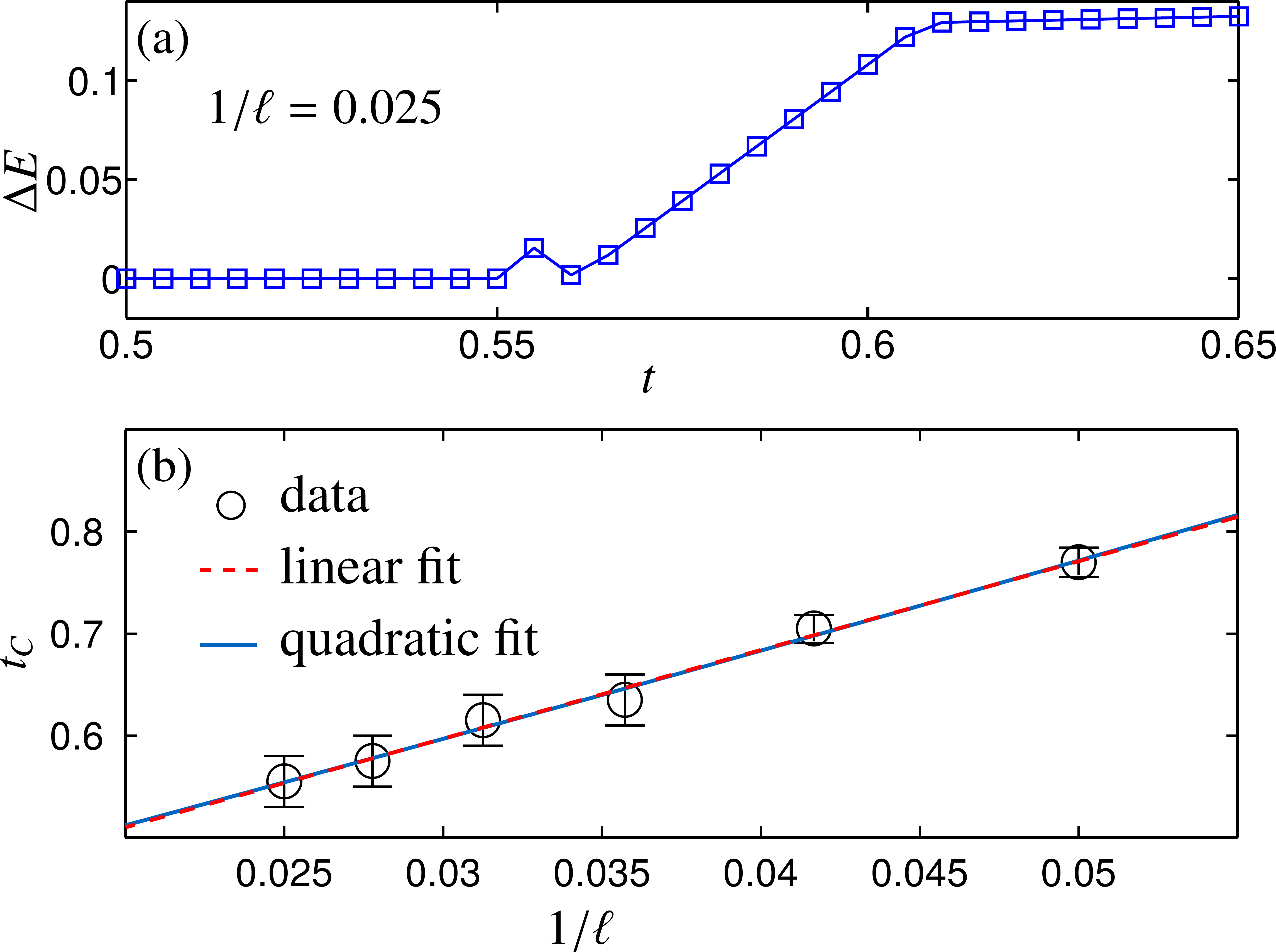}}
\caption{{(a) The gap from the ground state to the first ecited state as a function of the hopping amplitude $t$ for a fixed system size and fermion parity. To distinguish exact degeneracy (in the gapped phase) from a small gap (in the critical phase in the vicinity of $t_c$), highly accurate DMRG computations were performed by retaining up to 1000 states.} (b)The extrapolation of $t_c$ for the C-IC transition gives $t_c\approx 0.35$, setting $g=-1$.\label{fig:tc}}
\end{figure}

 We then determine $t_C$ (for $g=-1$ ) from our DMRG data by finding the largest value of $t$ at which there is an exact 
 4-fold ground state degeneracy (exact 2-fold degeneracy with, say, even fermion parity) for $L=8N$ and extrapolating this value of $t$ to large $\ell=L/2$. The results are shown in 
 Fig.~\ref{fig:tc}. We plot the value of $t_C$, where this exact degeneracy splits as a function of $1/\ell$. Linear ($a/\ell+t_C$) and quadratic ($a/\ell+b/\ell^2+t_C$) fits to $1/\ell$ give the same result for the extrapolation within the error bar of $0.025$ and we thus find $t_C/g=-0.35\pm 0.025$ ($g/t_C=-2.86$).

\section{Attractive Interactions and the  tricritical Ising point: $g>0$}
\label{ec:review}
The system for $g>0$ was studied in Ref.~\onlinecite{Rahmani2015}, where it was shown to realize the tricritical Ising model, a supersymmetric CFT with central charge $c=7/10$, at a critical value of $g/t$, which separates the $c=1/2$ Ising phase (at weak coupling) from a doubly-degenerate gapped phase (at strong  coupling). While the primary focus of this paper is on repulsive interactions, in this section, we review some of the salient results of Ref.~\onlinecite{Rahmani2015} for completeness. We also analyze a first-order phase transition that occurs between the symmetry broken phases at $g/t= \infty$ (fixed $g>0$ and $t=0$), which has not been discussed elsewhere.

The arguments of Sec.~\ref{sec:weak}, supporting a critical Ising phase around $g=0$, were independent of the sign of $g$. Therefore, the $c=1/2$ Ising phase is expected to  extend to a finite value of interaction strength $g$ also for attractive interactions. As we argued in Sec.~\ref{sec:strong}, unlike the 4-fold degeneracy of the strong-coupling limit gapped phase for $g<0$, the gapped phase at strong coupling is doubly degenerate in the case of $g>0$. A priori, it is not obvious that there is only one phase transition between these two strong- and weak-coupling phases. If this is the case, however, it is well known from the theory of an Ising model with vacancies~\cite{Cardy}, that the most natural phase transition between the critical Ising phase and a doubly degenerate gapped phase is the TCI CFT. 
This CFT is of great interest as it provides a rare example of emergent supersymmetry in condensed matter physics. 
\begin{figure}[t]
\centerline{\includegraphics[width=80mm]{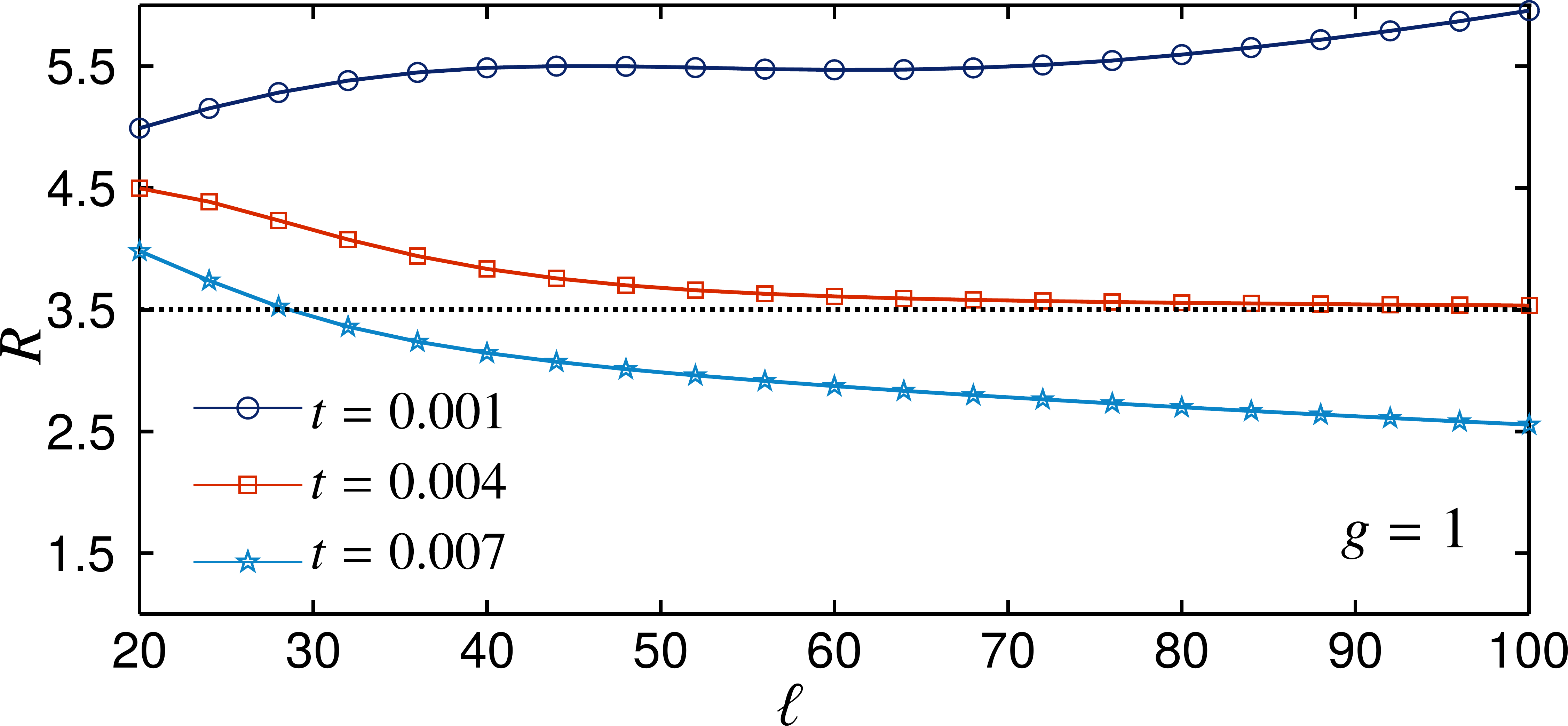}}
\caption{The gap ratio $R$ as a function of system size {computed with DMRG (keeping 600 states)} . A plateau emerges at $t/g\approx 0.004$ at the expected value of $7/2$ for the TCI CFT.}
\label{ratios}
\end{figure}

It turns out that the doubly degenerate gapped phase has a very large correlation length. Therefore, numerical verification of the above-mentioned scenario is exceedingly difficult. As showed in Ref.~\onlinecite{Rahmani2015}, however, universal ratios in energy gaps provide a powerful method of verifying the scenario and identifying the transition point corresponding to the TCI CFT. Our primary diagnostic is the ratio
\begin{equation}
R={E^{\rm odd}_{0}-E^{\rm even}_{0}\over E^{\rm even}_{1}-E^{\rm even}_{0}},
\end{equation}
where the superscripts even and odd indicate the fermion parity, $E_0$ ($E_1$) is the ground (first excited) state energy in the parity sector, and all energies are computed with antiperiodic boundary conditions. The finite-size spectra of CFTs can be derived from the scaling dimensions of their primary operators. In particular, for the TCI CFT we have
\begin{equation}
R_{\rm TCI}=7/2.
\end{equation}
As shown in Fig.~\ref{ratios}, the numerically computed value of $R$ plateaus at $t/g=0.004$ as a function of system size precisely at $R=7/2$ predicted for the TCI CFT.

Several other universal ratios support the presence of a transition between the Ising phase and the gapped phase through the TCI CFT. Unlike the $c=1/2$ Ising model, the low-energy fermionic excitations of the TCI CFT are not regular free Majorana fermions. This indicates nontrivial exponents for the Green's function $\langle \gamma(x,t)\gamma(0,0)\rangle$. In particular, the equal-time Green's function decays as $1\over x^{7/5}$ for the TCI CFT as opposed to $1\over x$ for the Ising case, as was verified in Ref.~\onlinecite{Rahmani2015}.


{\section{Self-consistent mean field theory and topological classification of the gapped phases}\label{sec:dimer}
In this section, we extend the mean-field-like picture of Sec.~\ref{sec:strong}, which provided a qualitative description of the strong-coupling limit, to a more systematic self-consistent mean-filed treatment. As argued in Sec.~\ref{sec:repul}, the Lifshitz transition can be understood in terms of the generation of an effective third-neighbor hopping in the renormalization group, which can renormalize the dispersion relation. Such terms are also naturally generated in the mean-filed decomposition, and as we show, the self-consistent mean field theory can provide an accurate description of this transition. We also provide a topological classification of the more general dimerized model defined in Eq.~\eqref{H'}.
}

Many 1D systems, such as  polyacetylene, are known to spontaneously
dimerize. In our present model, as realized by the 1D chain of vortices,
dimerization may also occur, leading to a model with explicitly broken
translational symmetry. Whether or not such a dimerization occurs  and its amplitude will depend on the details of the vortex lattice physics, specifically the intervortex interactions and vortex pinning. We do not attempt to specify the conditions under which dimerization may take
place or calculate its strength. Rather, we explore in this Section the phases of the dimerized
model and discuss their topological properties.

The relevant Hamiltonian (\ref{H'}) can be conveniently rewritten in terms
of new Majorana operators $\alpha_j=\gamma_{2j}$ and $\beta_j=
\gamma_{2j+1}$ as $H=H_0+H'$ with 
\bea
H_0&=&i\sum_j\left(t_1\alpha_j\beta_j+ t_2
  \beta_j\alpha_{j+1}\right) \label{d1} \\
H'&=&\sum_j\left(g_1 \alpha_j\beta_j\alpha_{j+1}\beta_{j+1} +
g_2\beta_j\alpha_{j+1}\beta_{j+1}\alpha_{j+2}\right). \label{d2}
\eea
We continue assuming that $t_1$ and $t_2$ are positive but consider
either sign of $g_1$ and $g_2$.
Although the original translation symmetry by one Majorana site is
broken the model obeys the antiunitary time-reversal symmetry ${\cal
  T}$ generated by $(\alpha_j,\beta_j)\to(\alpha_j,-\beta_j)$ and $i\to -i$. 
We note that in the language of previous Sections this corresponds to $\gamma_R\leftrightarrow\gamma_L$  and $i\to -i$, which is the proper form of  time-reversal in the relativistic field theory. 
In the absence of interactions it also obeys the antiunitary particle-hole
duality ${\cal C}$
generated by $(\alpha_j,\beta_j)\to(\alpha_j,\beta_j)$ and $i\to -i$
which maps $H_0\to -H_0$. This puts the noninteracting Hamiltonian
$H_0$ into the
BDI class under the Altland-Zirnbauer classification \cite{Altland1997}. In 1D its gapped 
topological phases are therefore classified by an integer invariant
$\nu$ \cite{Ryu2010}. If we define $\nu_{\alpha/\beta}$ as the number of unpaired
MZMs of type $\alpha/\beta$ bound to the left end of
the chain with open boundary conditions then the invariant
$\nu$ coincides with $\nu_\alpha-\nu_\beta$. Note that the two types of Majorana modes are distinguishable because they transform as even and odd, respectively, under ${\cal T}$. Also, because there is the same number of $\alpha$'s and $\beta$'s in the system for each $\alpha$ bound to the left edge there must be one $\beta$ bound to the right edge and vice versa.   

In their seminal work Fidkowski and Kitaev \cite{Fidkowski2010,Fidkowski2011}
showed that when interactions preserving ${\cal T}$ are added to such
a Hamiltonian, which is the case here, its integer classification is changed to the Z$_8$
classification: phases characterized by invariants $\nu$ and $\nu+8$ become
indistinguishable.

It is important to note that symmetry ${\cal T}$ of $H_0$ hinges on the absence of the second neighbor tunneling  because terms such as $i\alpha_j\alpha_{j+1}$ would clearly break  ${\cal T}$. In a physical system this may be realized to a good approximation due to the exponential decay of the MZM wavefunctions. Alternately, as demonstrated in Ref.\   \cite{Chiu2015b,Pikulin2015}, the  ${\cal T}$ symmetry  can be implemented exactly when the chain is realized in a system of alternating vortices and antivortices in the surface of an STI with the chemical potential tuned to the neutrality point.  When ${\cal T}$ is broken the noninteracting system is in symmetry class D and its classification in 1D is Z$_2$, with or
without interactions. Physically, then, only phases with even and odd index $\nu$ are distinct.

The Hamiltonian $H$ contains 3 dimensionless parameters. A detailed
analysis of this 3-dimensional parameter space using DMRG and exact
diagonalization would require a lot of computer time and we leave it to future
studies. Here, we perform a survey of its gapped phases using the
mean-field theory. Although such MF theories often fail to accurately
capture the physics of systems in low dimensions we expect the
description of gapped phases to be qualitatively correct over part of the phase diagram 
(although not the nature of the critical points). In some parameter ranges, we
find a good agreement between the MF results and the more accurate
DMRG calculations. 

To proceed we perform a MF decoupling of the interaction term $H'$ in
all channels respecting the symmetries of $H$. This leads to the MF
Hamiltonian of the form
\be
H_{\rm MF}=i\sum_j\left(\tau_1\alpha_j\beta_j+ \tau_2
  \beta_j\alpha_{j+1} +\tau'_1 \alpha_j\beta_{j+1} + \tau'_2
  \beta_j\alpha_{j+2}\right), \label{d3}
\ee
where $\tau$'s are the MF parameters.
{Note that, in the translationally invariant case, $\tau_i'$ corresponds to
the third neighbor hopping $t'$ introduced in Eq. \eqref{eq:3rd}.}
 These are determined based on
the requirement that the ground state $|\Psi_{\rm MF}\rangle$ of
$H_{\rm MF}$ minimizes the expectation value $\langle H\rangle_{\rm
  MF}=\langle\Psi_{\rm MF}|H|\Psi_{\rm MF}\rangle$. This leads to a
system of self-consistent MF equations for parameters $\tau$ that read
\bea 
\tau_1&=&t_1+{1\over N}\sum_{k>0}\left(2g_1{\partial
    E_k\over\partial\tau_2}+g_2{\partial E_k\over\partial\tau'_1}\right)  \label{d4}\\
\tau'_1&=&{1\over N}\sum_{k>0}g_1{\partial E_k\over\partial\tau_1}. \label{d5}
\eea
and similar equations for $\tau_2$ and $\tau'_2$ obtained by interchanging
all indices $1\leftrightarrow 2$. Here, $N$ denotes the number of unit
cells in the system,  $k$ extends over one half of the Brillouin zone $(-\pi,\pi)$, which is now half the size of the BZ used in Sec. III because the unit cell has been doubled due to dimerization.
\be\label{d6}
E_k=4\sqrt{\left(\tau_+\sin{k\over 2}+\tau'_+\sin{3k\over 2}\right)^2+
\left(\tau_-\cos{k\over 2}+\tau'_-\cos{3k\over 2}\right)^2},
\ee
is the spectrum of excitations of $H_{\rm MF}$ and $\tau_\pm=(\tau_1\pm\tau_2)/2$.

The physics of the MF approximation has a simple intuitive
interpretation. In the noninteracting limit $H_{\rm MF}$ coincides
with $H_0$. The interaction terms present in $H'$ are seen to
renormalize the nearest neighbor hopping terms $t_1$ and $t_2$ via
Eq.\ (\ref{d4}) and generate third-neighbor hoppings through  Eq.\ (\ref{d5}).
Second neighbor hoppings would violate ${\cal T}$ and are therefore
not generated. For weak interactions it is easy to see that the MF
equations imply an increase in  hoppings with increasing $|g|$ for $g$
positive but decrease when $g$ is negative. This observation provides
an intuitive explanation for the qualitatively different behavior of
the model depending on the sign of $g$ found in the previous
Sections.  We also note that the MF theory becomes exact in the strong coupling limit when, say $g_2=t_2=0$ and for positive $g_1$. {[Notice that when $g_2=t_2=0$, the Hamiltonian reduces to Eq.~\eqref{Hdir2}, in which the operator $\hat{n}_j=(\hat{p}_j+1)/2$ is conserved. The MF approximation linearizes the fluctuations of $\hat{n}_j$ and is exact if no such fluctuations are present.] }In this limit Eqs.\ (\ref{d4},\ref{d5}) reproduce the ``ferromagnetic'' ground state discussed in Sec.\ II.B. For negative $g_1$ the expected ``antiferromagnetic'' ground state breaks the symmetry under translation by two Majorana sites built into $H_{\rm MF}$ and is therefore not captured by this MF theory, although a more general MF theory could be constructed to describe this state.

The MF analysis proceeds in two steps. First, for given parameters
$(t_1,t_2,g_1,g_2)$ we find the  MF Hamiltonian $H_{\rm MF}$ that best
approximates $H$ by
solving Eqs.\ (\ref{d4},\ref{d5}) to find MF parameters
$(\tau_1,\tau_2,\tau'_1,\tau'_2)$. This step must be performed
numerically. Second, we determine the topological phase characterizing
the ground state  of $H_{\rm MF}$ with these parameters. 

Because of
the large parameter space involved in the analysis it is instructive
to start with the second step and enumerate the possible topological
phases of the noninteracting MF Hamiltonian (\ref{d3}). This is
most easily done by studying its spectrum of excitations (\ref{d6}). We
adopt a reasonable assumption that the gapped regions represent distinct
topological phases with phase transitions marked by gap closings. We
furthermore adopt $\tau_+$ as our unit of energy and work with 3
dimensionless coupling parameters
\be
r={\tau'_+\over \tau_+}, \ \ \ s={\tau_-\over \tau_+}, \ \ \ s'={\tau'_-\over \tau_+}.
\ee
 For $E_k$ to be gapless both 
brackets under the square root in Eq.\ (\ref{d6}) must separately
vanish for the same momentum $k$. This imposes two
conditions on three  parameters $(r,s,s')$ and momentum $k$, leading to
the conclusion that phase transitions occur at a set of two-dimensional
surfaces in the 3-parameter space of $H_{\rm MF}$.
 We expect this result to remain valid beyond the MF theory. When the translation symmetry is broken, a mass term becomes allowed in the low-energy effective Hamiltonian (\ref{heff1}). The mass $m$, then, is a function of 3 dimensionless parameters that can be constructed from couplings $(t_1,t_2,g_1,g_2)$. Phase transitions between massive phases correspond to $m=0$ which imposes a single condition on 3 dimensionless parameters, leading to the same conclusion as above. In the Luttinger Liquid + Ising phase 
a mass term, $im\gamma_R\gamma_L$ in the Ising sector and a pairing term $\Delta (\psi_R\psi_L+h.c.)$ 
are allowed by symmetry in the dimerized model.  A single condition is enough to make either $m$ or $\Delta $ vanish 
corresponding to either a massless Ising model or a massless Luttinger Liquid. 

 The first bracket in the spectrum in Eq.\ (\ref{d6}) vanishes (i) for
$k=0$ and all values of $r$ or, (ii) for $k=k_0$ with $r$ given by 
\be\label{d8}
r={1\over 4\sin^2{(k_0/2)}-3}.
\ee
This solution exists only when $r\leq -{1\over 3}$ or  $r\geq 1$. Now
we must find for which values of $s$ and $s'$ the second bracket
vanishes at these values of $k$.

\begin{figure}[t]
\includegraphics[width = 8.0cm]{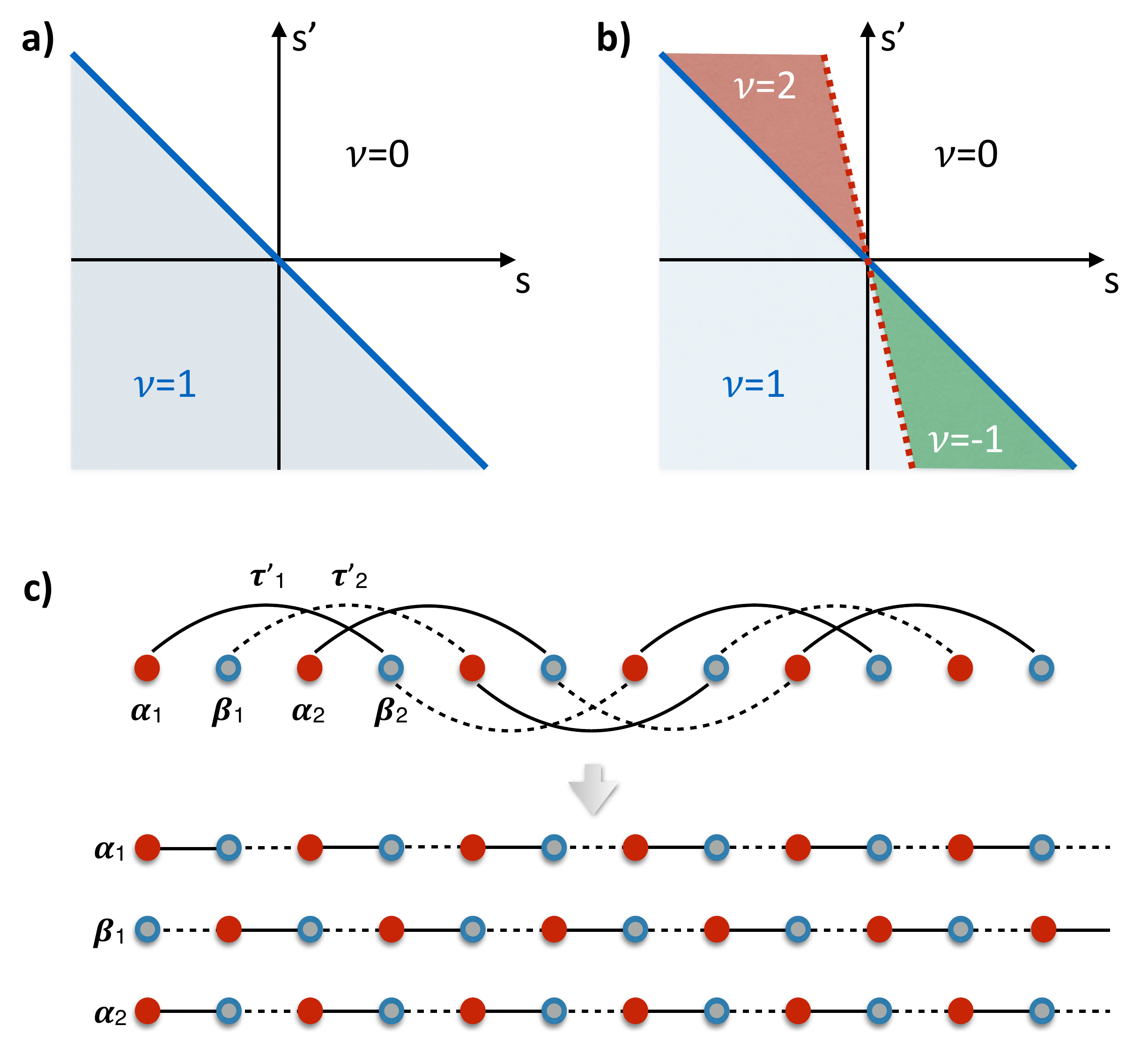}
\caption{Phase diagram of the mean-field Hamiltonian $H_{\rm MF}$. Topological phases for a) $r\in (-{1\over 3},1)$ and b) $r<-{1\over 3}$. The dashed phase boundary coincides with the solid line for $r=-{1\over 3}$, then rotates clockwise with increasing $|r|$, eventually reaching slope $+{1\over 2}$ as $r\to -\infty$.  c) A chain with only third nn hoppings is equivalent to three decoupled Kitaev chains. When $\tau_1'<\tau_2'$ the top and the bottom Kitaev  chain will be in the topological phase while the middle chain will be trivial, leading to the invariant $\nu=2$. In the opposite case the roles switch and we obtain $\nu=-1$. 
}\label{figMF1}
\end{figure}

Consider first $r\in (-{1\over
  3},1)$. The second bracket vanishes for $k=0$ when $s'=-s$. There
is, therefore, a single phase transition in the $s$-$s'$ plane
indicated in Fig.\ \ref{figMF1}a. The transition takes place between a topological
phase with $\nu=1$ and a trivial $\nu=0$ phase. This can be deduced by
considering the limit $r=s'=0$ in which $H_{\rm MF}$ coincides with
the Hamiltonian of the Kitaev chain with nn hopping whose classification is well
known. Adiabatic continuity then insures that the invariant $\nu$
remains unchanged unless we cross a phase boundary, hence the
identification of phases in Fig.\ \ref{figMF1}a.

Now consider $r\leq -{1\over 3}$. In addition to the $k=0$ solution we
now have a solution at $k=k_0$ with $k_0$ given by Eq.\
(\ref{d8}). The first solution implies a phase transition at $s'=-s$
as before while the second implies another phase transition line given
by
\be\label{d9}
s'=\left({r\over 2r+1}\right)s.
\ee
The phase diagram is shown in Fig.\ \ref{figMF1}b. In addition to the $\nu=0,1$
phases present before two new topological phases with $\nu=-1,2$
appear. The identification of these phases follows from observing that
the gap closing described by Eq.\ (\ref{d9})
(dashed line in Fig. \ref{figMF1}b) involves two Majorana modes and one
thus expects $\nu$ to change by $\pm 2$ across this line. Alternately,
one can consider the limit $\tau_\pm\to 0$ (corresponding $|r|,|s'|\gg |s|$)
in which the third nn hoppings dominate. In this limit the system
breaks up into 3 weakly coupled Kitaev chains as illustrated in Fig.\ \ref{figMF1}c. Two of these chains have an
$\alpha$ operator on their left edge and one has a $\beta$ operator
there. The two possible phases in this limit are thus characterized by
$\nu=2,-1$. To verify this phase assignment we have also explicitly computed the index $\nu$ in a system with periodic boundary conditions and found agreement with Figs.\ \ref{figMF1}a,b. Similar analysis applies to the parameter region $r>1$ with the same 4 distinct phases possible but we find that this regime is never reached in the solution of MF equations  (\ref{d4},\ref{d5}) and is therefore not relevant to our original interacting problem.

\begin{figure}[t]
\includegraphics[width = 8.0cm]{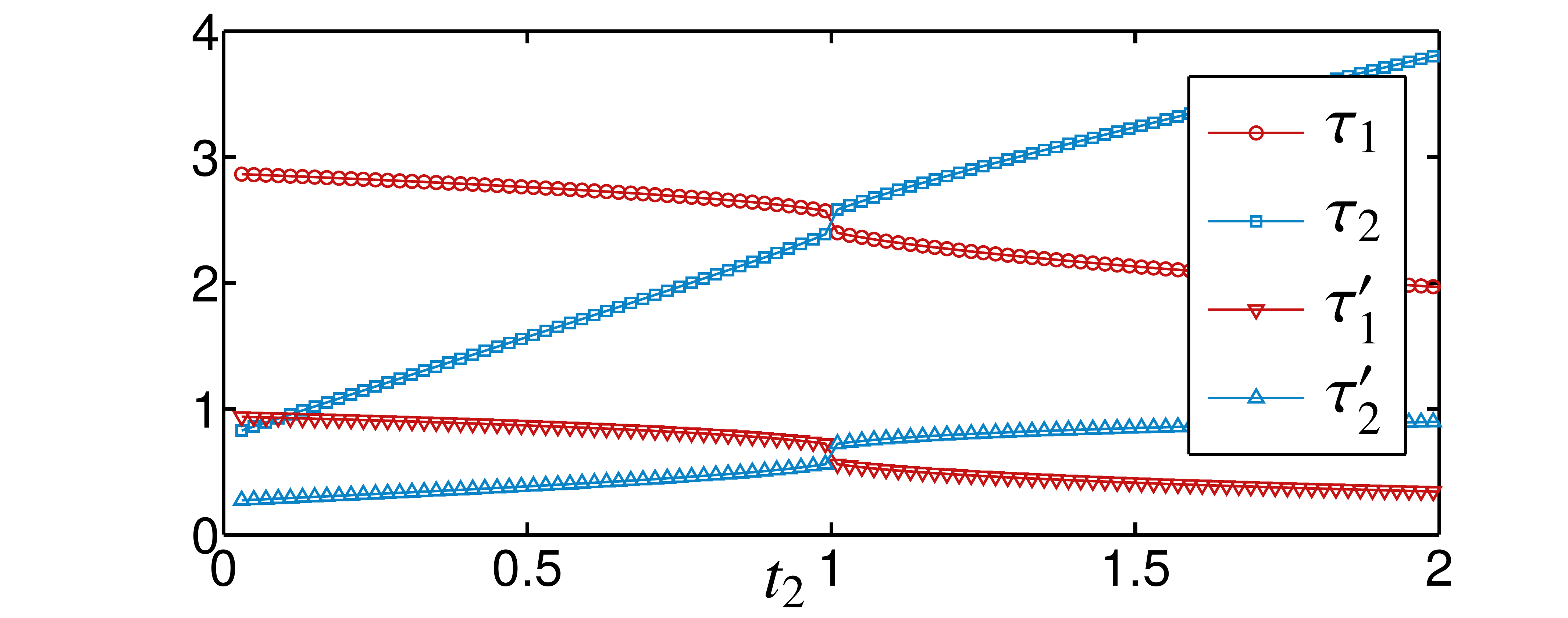}
\caption{Mean field parameters as a function of $t_2$ for fixed $t_1=g_1=g_2=1$. A jump at $t_2=1$ indicates the first order transition between topological phases with $\nu=0$ and $\nu=1$. 
}\label{figMF2}
\end{figure}

We now proceed to analyze the MF equations  (\ref{d4},\ref{d5}). For concreteness and simplicity we set $g_1=g_2=g$, although it is no more difficult to analyze the general case. For $g>0$ we find that only two phases indicated in Fig.\ \ref{figMF1}a appear in the MF theory and the transition occurs at $t_1=t_2$, as could be expected on the basis of symmetry when $g_1=g_2$. The transition is second order when $g=0$ but becomes weakly first order for any $g\neq 0$. We have seen in Sec. II that the actual transition in the interacting problem remains second order up to large values of $g$ so this is an artifact of the MF approximation. We note, however, that for weak coupling the excitation gap at the transition point is exponentially small ($\sim e^{-t/g}$) so the MF theory provides at least a qualitatively correct description of the transition at weak coupling (a spectrum with an exponentially small gap will be, for practical purposes, indistinguishable from a truly gapless spectrum). Fig.\ \ref{figMF2} shows the discontinuity in the MF solution for $\tau_1$ and $\tau_2$ for an intermediate coupling strength $g=1$, clearly indicating the first order transition. For larger $g$ the discontinuity becomes more prominent but no new phases appear.  
 
For repulsive interactions $g<0$ the behavior is qualitatively different. Consider first the high-symmetry situation $t_1=t_2=t$. Numerical solution of the MF equations  (\ref{d4},\ref{d5}) indicates that for $g<0$ the effective nn hopping $\tau$ decreases with increasing $|g|$ from its initial value $\tau=t$ at $g=0$. At the same time $\tau'$ also decreases, starting from zero and moving towards the negative values. When the velocity $v=2(\tau+3\tau')$ vanishes, a Lifshitz transition occurs with additional gapless branches of excitations  making their appearance at $k=k_0$. To make contact with our work in Sec.\ III we plot in  Fig.\ 4 the velocity $v$ extracted from the MF solution as a function of $t$ for $g=-1$. We observe that it vanishes at $t_c\simeq 3.41$, reasonably close to the value 3.52 found from DMRG for a fully interacting problem. 

Now consider the gapped phases reached by perturbing the system away from the $t_1=t_2$ symmetry line for $g<0$. When $t>t_c$ such perturbations only give rise to $\nu=0,1$ phases indicated in the phase diagram Fig.\ \ref{figMF1}a. This is suggested by the solution of the MF equations and we have also verified this by an exact numerical diagonalization of the full interacting Hamiltonian (\ref{d1},\ref{d2}).  For $t<t_c$, MF theory suggests that all four distinct phases with $\nu=-1,0,1,2$ shown in Fig.\ \ref{figMF1}b can be reached. This is in accord with the intuition that multiple species of low-energy gapless Majorana modes, once gapped by symmetry breaking perturbations, should give rise to phases with multiple MZMs bound to the edges of the chain. On the other hand, as indicated in Fig.\ \ref{figMF2}(c) we expect the new $\nu=-1,2$ phases to be reached only when the effective third nn hoppings dominate. In the model defined by Eqs.\ (\ref{d1},\ref{d2}) this will occur for relatively strong interaction strength because the third nn hoppings are absent when $g=0$. In this regime we expect the MF theory to be at best qualitatively correct and our limited search using the exact diagonalization method did not find conclusive evidence for these phases. The $\nu=-1,2$ phases can be stabilized by adding the third nn hoppings to the interacting Hamiltonian. They are allowed by symmetries and will generically be present in the physical system. We leave the detailed investigation of the resulting phases for future studies.

We close this Section by noting that in the parameter regions where only $\nu=0,1$ phases are present spontaneous dimerization in the geometry with open boundary conditions may favor the phase without MZMs. This can be seen by thinking about the strong coupling ground states of the model visualized in Figs.\ \ref{g>} and \ref{g<}. Clearly, the state with all MZMs combined into Dirac fermions will be lower in energy than the state with two unpaired MZMs at the edges. If there is a region in the parameter space where only the $\nu=-1,2$ phases exist, presumably, by the same argument the $\nu=-1$ phase will be energetically favored over the $\nu=2$ phase.

\section{experimental realizations and signatures}

As argued previously in Refs.\ \cite{Chiu2015,Chiu2015b} all the ingredients are now in place to start experimentally exploring systems of strongly interacting Majoranas including the simple 1D system discussed in this work. The most promising physical platform is the superconducting surface of an STI where MZMs are bound in the cores of Abrikosov or Josephson vortices \cite{Fu2008}. Experimentally, superconducting order has been induced in such surfaces by multiple groups and in several different STI materials \cite{Kor11, Sac11, FQu12, Wil12, Cho13, YXu14, Zha14, PXu14,Harlingen2014}. The ability to tune the  chemical potential  to the vicinity of the Dirac point, required to bring in the regime of  strong interactions, has also been demonstrated \cite{Cho13, YXu14, Zha14}. Recently, individual vortices have been imaged in these systems \cite{PXu14} and spectroscopic evidence indicative of MZMs in the cores of  vortices has been reported \cite{PXu15}. 1D structures such as those envisioned in this work most naturally arise in Josephson junctions built on the STI surface. When magnetic field is applied perpendicular to the surface of the STI a line of Josephson vortices is known to form inside the junction. Such Josephson vortices carry MZMs and their spacing can be conveniently controlled by the magnetic field amplitude. Evidence suggestive of MZMs in such devices has recently been reported \cite{Harlingen2015}. 

The chain of 1D interacting Majoranas could be realized in other physical systems that are known to host MZMs \cite{Alicea2012,Beenakker2012,Elliott2015}. This includes semiconductor quantum wires with strong spin-orbit coupling and the edge of a 2D topological insulator. If these are coupled to a periodic structure made of alternating superconducting and magnetic regions then  MZMs are expected to form at the boundaries between the corresponding domains. In the presence of interactions the physics of such MZMs will  be described by Hamiltonian (\ref{eq:hamil}) studied in this work.

Scanning tunneling microscopy (STM) can provide valuable experimental signatures of the phases and phase transitions of this system when the model is realized by  MZMs bound to vortex cores or other structures as discussed above. Importantly, the experimental ability to tunnel into a Josephson junction region has previously been demonstrated \cite{Pill10,Seu08}, both with STM and with planar contact tunneling. Below we address the characteristic signatures of various phases and phase transitions present in our model that are observable through the tunneling conductance.

The tunneling current between the sample and the normal tip goes as $\langle I\rangle\propto G_R(-eV)$, where the retarded Green's function is \cite{mahan1}
 \be G_R(\omega)=-i\int_0^\infty dt e^{i\omega t}\langle[\gamma_j(t)\psi_0(t),\gamma_j(0)\psi^\dagger_0(0)]\rangle.\ee
Here the operator $\psi_0$ annihilates an electron at the tip. Due to a factorization of the time-ordered Green's function into tip and sample correlators, the tunneling current captures the behavior of the temporal correlation functions of the Majorana chain, which are closely related to the spatial (equal-time) correlators at low energies due to the emergent Lorentz invariance.

Several experimental signatures follow from this. For example, in the Ising phase, the Green's function decays with the characteristic free-fermion exponent as $x^{-1}$ and we find $I_{\rm Ising}\propto V$. At the tricritical point, the leading (with the smallest scaling dimension) fermionic operator has scaling dimension $7/10$. Therefore, as shown in Ref. \cite{Rahmani2015}, the Green's function decays as $x^{-7/5}$ and the tunneling current goes as 
\be
I_{\rm TCI}\propto {\rm sgn}(V)|V|^{7/5}.
\ee
 In the 2-fold degenerate gapped (massive) phase, the power-law dependence of the tunneling current on $V$, which occurs in massless Ising and tricritical phases, becomes an exponential dependence.
%
%
%

On the negative-$g$ side, the Lifshitz transition is characterized by a dynamical exponent $z=3$, which changes the constant density of states of the Ising phase to a density of states at energy $\epsilon$ proportional to $\epsilon^{-2/3}$. This leads to
 \be
I_{\rm Lifshitz}\propto |V|^{1/3}.
\ee
 Interestingly, at the Lifshitz transition the tunneling conductance ${d I\over dV}$ diverges for small $V$. In the Ising+LL phase, the Green's function decays as $x^{-1}$ to leading order, with subleading corrections of the form $x^{-(K+1/K)/2}$ [see. Eq.~\eqref{eq:corr} and Fig.~\ref{fit}(b)], indicating a tunneling current linear in $V$ with subleading corrections (for small $V$) scaling as $V^{(K+1/K)/2}$.

Although the contribution of the LL sector to the tunnelling current in the Ising+LL phase is subdominant in $V$, it may be 
possible to see it clearly by doing STM near the end of a Majorana chain. There we expect a Luttinger-liquid contribution that oscillates spatially at wave-vector $2k_0$ while decaying with a $K$-dependent 
power law with the distance from the end of the chain~\cite{Eggert2000}. Observing the values of $k_0=\pi/4$ and $K=1/4$, followed by a breakdown of the power-law dependence of the tunneling current on $V$, when entering the 4-fold degenerate gapped phase, can then signal the C-IC transition.

\section{conclusions}

The spinful and spinless variants of the Hubbard model in 1D, which respectively have on-site and nearest-neighbor interactions, are two widely studied canonical models of interacting Dirac fermions. Here, we studied a third canonical model, namely, the minimal 1D model of interacting Majorana zero modes. These Majorana degrees of freedom have Hermitian creation operators and are effectively half of a (complex) Dirac fermion. In this case, the minimal interaction involves four sites.

In light of the focused experimental efforts on realizing and manipulating Majorana zero modes as emergent particles in solid-state devices, the behavior of interacting many-body Majorana systems is of great theoretical as well as experimental interest. 
We argued that ingredients necessary to begin experimental investigations of such systems with strong interactions are presently in place. Such investigations could naturally start with model defined in Eq.\  (\ref {eq:hamil}), this being the simplest interacting 1D Hamiltonian one can construct with these degrees of freedom.

Using a combination of analytical techniques based on field-theory and RG, mean-field calculations, and numerical DMRG studies, we determined the full phase diagram of this novel strongly correlated system.
The physics that emerges from this simplest model of interacting Majoranas is extremely rich and complex, revealing novel phases and phase transitions that are not present in the well studied Dirac counterparts. As previously shown in Ref.~\cite{Rahmani2015}, this model provides one of the few examples of emergent space-time supersymmetry for attractive interactions. In the present  paper we extended the analysis of the model to the case of repulsive interactions. We found a novel $z=3$ quantum critical point, at which a Lifshitz transition occurs changing the topology of the Fermi surface. On the weak-coupling (strong-coupling) side of the Lifshitz transition, we have one (three) species of low-energy Majoranas. In the Ising phase, the Majoranas are free, while on the strong-coupling side (the Ising+LL phase), only one species remain free. The other two species combine into interacting Dirac fermions forming a Luttinger liquid, with an emergent Fermi momentum and particle number (despite the fact that particle number conservation is not a symmetry of the Hamiltonian).

At even larger repulsive interactions, the Ising+LL phase undergoes a transition to a 4-fold degenerate gapped phase. The nature of this novel phase transition, which fully gaps out a $c=3/2$ CFT is reminiscent of the commensurate-incommensurate transition. As the emergent 
Fermi momentum varies with interactions, a term in the Hamiltonian, which generically has fast oscillations, becomes nonoscillatory at a particular commensurate wave vector and drives the transition.

When the translation symmetry of the chain is explicitly broken by dimerization, as often happens in 1D systems, additional phases can appear. We performed a brief survey of these dimerized phases using mean-field theory and found four distinct gapped phases characterized by an integer topological invariant $\nu$. The latter takes values $-1,0,1,2$, and can be interpreted as the number of unpaired Majorana zero modes bound to the edge of the chain in the geometry with open boundary conditions.   

Our work extends the space of canonical 1D models of interacting fermions from the Hubbard chain and its spinless variant to a third simple and experimentally relevant model describing the most natural interacting system composed of Majorana zero modes, revealing a plethora of novel phases and phase transitions. 

When this paper was almost complete, 
we became aware of~\cite{Milsted2015}, which contains  related results.

\begin{acknowledgments}
This work was supported by NSERC (IA, MF, and AR), CIfAR (IA and MF), Max Planck-UBC Centre for Quantum Materials (IA, MF, and AR) and China Scholarship Council (XZ). 
\end{acknowledgments}

\appendix{}
\section{Weak coupling}
\label{ap:1}

 While it is possible to diagonalize the noninteracting 
limit of Eq. (\ref{HDir}) by a Bogliubov-DeGennes transformation, it is much simpler to Fourier transform the 
Majorana operators. For a chain of $L$ Majorana operators, $\gamma_1, \gamma_2, \ldots \gamma_L$:
\be \gamma_j=\sqrt{2\over L}\sum_ke^{-ikj}\gamma (k), \ \  (-\pi \leq k\leq \pi )\ee
where $k=2\pi n/L$ with periodic boundary conditions or 
$k=2\pi (n+1/2)/L$ with anti-periodic boundary conditions. Inverting, we obtain:
\be \gamma (k)=\sqrt{1\over 2L}\sum_ke^{ikj}\gamma_j.\ee
This implies:
\be \{\gamma (k),\gamma (k')\}=\delta_{k,-k'}, \ \  \gamma (-k)=\gamma^\dagger (k).\ee
The Fourier transformed Hamiltonian is simply:
\be H=2t\sum_k \gamma (-k)\gamma (k)\sin k.\ee
This is already diagonalized and we see that we should identify $\gamma (k)$ as an annihilation operator 
for $k>0$ and as a creation operator for $k<0$:
\be H_0=4t\sum_{0<k<\pi}\gamma^\dagger (k)\gamma (k)\sin k-{2t\over \sin (\pi /L)}.\ee
In order to treat interactions using the renormalization group we focus on the low energy excitations, 
occurring at $k$ near zero and $\pi$ where the dispersion relation becomes linear with slope $v=4t$. 
In Schroedinger representation, we may write:
\be \gamma_j(t)\approx 2\gamma_R(vt-j)+(-1)^j2\gamma_L(vt+j)\ee
where 
\bea \gamma_{R}(vt-j)&\equiv& 
\sqrt{1\over 2L}\sum_{0<k<\Lambda} \left[e^{-ik(vt-j)}\gamma (k)+e^{ik(vt-j)}\gamma^\dagger (k)\right] \\
\gamma_{L}(vt+j)&\equiv& \sqrt{1\over 2L}
\sum_{0<k<\Lambda} \left[e^{-ik(vt+j)}\gamma (\pi -k)+e^{ik(vt+j)}\gamma^\dagger (\pi -k)\right]. \nonumber 
\eea
Here $\Lambda$ is a momentum cut-off, $\Lambda \ll 1$. (We work in units where the lattice spacing is set to 1.) 
$\gamma_{R/L}(vt\mp j)$ is a relativistic right/left moving Majorana fermion field.

\section{Spectrum of the Ising+LL phase}
\label{ap:2}

 Let us first review the 
finite-size spectrum for noninteracting relativistic Dirac fermions with Fermi wave-vector $k_F$, with APBC. This 
depends on a number $f$, with $|f|\leq 1/2$,  which is the fractional part of $k_FL/(2\pi )$.  The energy to add $N_R$ fermions 
in the lowest energy states above the Fermi energy on the right branch is:
\be E={2\pi v\over L}\sum_{n=0}^{N_R-1}(n+1/2-f) = {2\pi v\over L}\left({1\over 2}N_R^2-fN_R\right).\label{ER}\ee
Taking $N_R$ to be a negative integer, this formula also gives the energy to create $N_R$ holes in the 
lowest energy states. Thus Eq. (\ref{ER}) gives the energy for the lowest excited state with charge $N_R$ relative 
to the charge of the ground state. After adding $N_R$ particles in the lowest energy states we can make 
arbitrary particle-hole excitations.  If $N_{nR}$ particles are raised by $n$ energy levels the energy cost is 
$(2\pi v/L)nN_{nR}$. So the energy for a general 
particle-hole excitation of right movers is
\be E_{R,ph}={2\pi v\over L}\sum_{n=1}^\infty N_{nR}n.\ee
The same formulas hold for left-moving excitations. Combining them, it is convenient to define:
\bea N&\equiv& N_R+N_L\nonumber \\
M&\equiv& N_R-N_L.\eea
Note that $N$ and $M$ must have the same parity:
\be N=M,\ \  (\hbox{mod}\ 2).\ee
The excitation energy for an arbitrary low energy state is:
\be \Delta E={2\pi v\over L}\left[ {1\over 4}(N-2f)^2+{1\over 4}M^2+\sum_{n=1}^\infty (N_{nR}+N_{nL})n\right].
\label{fssLL}\ee

These formulas carry over directly to the low energy excitations of the noninteracting Majorana system for $t'<-t/3$, 
with $k_F$ replaced by $k_0$. The analog of creating a right-moving hole corresponds to creating an 
excitation at $k$ slightly less than $-k_0$. 

In addition, we may create excitations in the Ising sector.
Adding $I_R$ particles to the lowest energy states at small positive $k$ and $I_L$ particles to the 
lowest energy states at $k$ slightly larger than $-\pi$ costs energy 
\be \Delta E_I={2\pi v_0\over L}\left[ {1\over 4}N_I^2+{1\over 4}M_I^2\right].\ee
Again
\bea N_I&=&I_R+I_L\nonumber \\
M_I&=&I_R-I_L\eea
and $N_I$ and $M_I$ have the same parity. The formula for particle-hole excitations is more complicated 
in this case although it is well known. We won't need it to analyze our DMRG data. It is convenient
to shift $N$ in Eq. (\ref{fssLL}) by twice  the integer part of $k_0L/2\pi$, yielding Eq. (\ref{fss}). 

\begin{figure}
\centerline{\includegraphics[width=80mm]{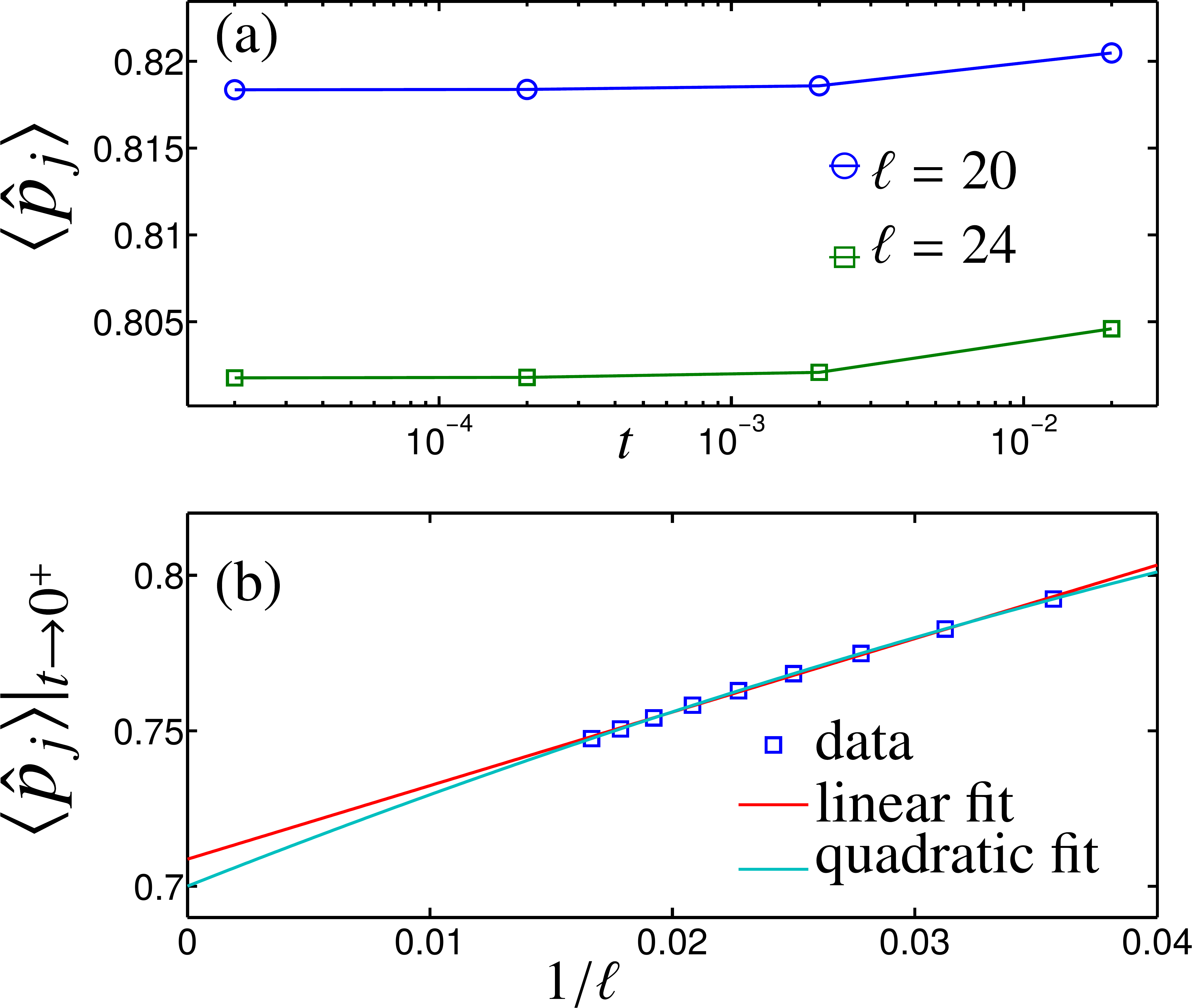}}
\caption{(a) The expectation value of $\hat p_j$, for PBC {(computed with DMRG and retaining 700 states)}, 
 goes to a nonzero value as $t\to 0^+$ for $g=1$ even in finite systems. As $\langle \hat p_j \rangle$ changes sign when $t\to-t$, this indicates a discontinuous jump in $\langle \hat p_j\rangle$. (b) Extrapolating $\langle \hat p_j\rangle|_{t\to 0^+}$ to the thermodynamic limit strongly suggests that the jump survives at $\ell\to \infty$.}
\label{fig:jump}
\end{figure}

\section{Numerical evidence for the first-order transition between symmetry-broken phases for $g>0$}\label{app3}
As discussed in Sec. II B, the first order transition corresponds to a splitting of the 4-fold degenerate ground states of Fig.\ 2 into  2-fold degenerate ground states plus 
a degenerate pair of excited states as $t$ is turned on for $g=1$. Depending on the sign of $t$ either the states with filled ($\langle\hat p_j\rangle=1$) or empty ($\langle\hat p_j\rangle=-1$) Dirac levels are favored 
(first and third or second and fourth states in Fig.\ 2). With PBC, the surviving 2-fold degenerate states have 
opposite fermion parity, as argued above. [See Eq. (\ref{eq:FF}.]
If we focus on the 
states with even fermion parity, 1st and 3rd in Fig.\ 2, then they are distinguished by $\langle \hat p_j\rangle$. For $t>0$ ($t<0$) the states with 
$\langle \hat p_j\rangle<0$ ($\langle \hat p_j\rangle>0$) occur. The transformation $\gamma_{2j}\to-\gamma_{2j}$, which leads to $t\to-t$, implies that $\langle \hat p_j\rangle(t)=-\langle \hat p_j\rangle(-t)$. The first-order transition should be signaled by a jump in $\langle \hat p_j\rangle$ at $t=0$.
 We found that this jump occurs even for finite systems. As seen in Fig.~\ref{fig:jump}(a), the value of $\langle p\rangle$ saturates as $t$  approaches $0^+$ (notice the logarithmic horizontal axis). In Fig.~\ref{fig:jump}(b), we extrapolate the value of $\langle \hat p_j\rangle$ for $t\to 0^+$ to $\ell\to\infty$ and find that it goes to $0.70\pm 0.01$, clearly indicating a jump as we cross $t=0$.

\bibliography{majorana.bib}

\end{document}